\documentclass[acmtog]{acmart}

\def\BibTeX{{\rm B\kern-.05em{\sc i\kern-.025em b}\kern-.08emT\kern-.1667em\lower.7ex\hbox{E}\kern-.125emX}}
    
\copyrightyear{2021}
\acmYear{2021}
\setcopyright{acmlicensed}
\acmConference[ICDCN 2022]{23rd International Conference on Distributed Computing and Networking}{January 4--7, 2022}{Delhi, India}
\acmBooktitle{23rd International Conference on Distributed Computing and Networking, January 4--7, 2022, Delhi, India}
\usepackage{subcaption}
\usepackage[utf8]{inputenc}
\usepackage{tikz}
\newtheorem{assumption}{Assumption}

\newtheorem{definition}{Definition}

\edef\qedrestoreat{\noexpand\catcode\lq\noexpand\@=\the\catcode\lq\@}%%
\catcode\lq\@=11

\ifx\protect\undefined\let\protect\relax\fi

\def\qed{\protect\@qed{$\qedsymbol$}}% suppressible printing of \qedsymbol
\def\QED{\protect\@qed{{\rm Q.E.D.}}}%  "quod   erat  demonstrandum"

\def\QEI{\protect\@qed{{\rm Q.E.I.}}}%  "quod   erat  inveniendum"

\def\QEF{\protect\@qed{{\rm Q.E.F.}}}%  "quod   erat  faciendum"

\def\Proof{\protect\@Proof}\def\endProof{\protect\@endProof}%

\def\qedsymbol{\square}% PLEASE NOTE: this is in the AMS symbols font.
\def\TheWordProof{\bf Proof\enskip}

\def\ProofFont{}

\newif\ifAutoQED\AutoQEDfalse

\newif\ifNumberResults
\ifx\theorem@style\undefined
   \NumberResultstrue
   \def\TheoremHeader#1#2#3{\bf #1\ifNumberResults\ #2\unskip\fi#3}
   \def\TheoremFont{\it}
\fi

\def\TheoremsAsCommands{%
  \def\TheoremFont{}% suppress italicised theorems.
  \def\begin@theorem##1{%
     \par\removelastskip\smallskip
     \save@set@qed % enable \qed
     \noindent{##1}% the name
   }%
  \def\@endtheorem{\ifAutoQED\qed\fi\restore@qed}%
}%

\expandafter\let\csname ds@theorems-as-commands\endcsname\TheoremsAsCommands

\def\parag@pushright#1{{% set up
    \parfillskip=0pt            % so \par doesnt push \square to left
    \widowpenalty=10000         % so we dont break the page before \square
    \displaywidowpenalty=10000  % ditto
    \finalhyphendemerits=0      % TeXbook exercise 14.32
    \hbox@pushright             % this used to be incorporated
    #1%                         % the end-of-proof mark (or whatever)
    \par}}%                     % build paragraph with the above parameters

\def\hbox@pushright{% horizontal
    \unskip                     % remove previous space or glue
    \nobreak                    % don't break lines
    \hfil                       % ragged right if we spill over
    \penalty50                  % discouragement to do so
    \hskip.2em                  % ensure some space
    \null                       % anchor the following \hfill
    \hfill                      % push \square to right
}%

\def\vbox@pushright#1{\expandafter\message %  (2 Jan 1994)
  {QED.sty could be improved in this case (line \the\inputlineno): please ask}%
  \page@pushright{#1}}%

\newif\if@qed\@qedfalse
\def\save@set@qed{\check@pt@sty@v\let\saved@ifqed\if@qed\global\@qedtrue}%
\def\restore@qed{\global\let\if@qed\saved@ifqed}

\def\@Proof{%
   \par\removelastskip\smallskip\penalty700
   \save@set@qed
   \noindent\ProofFont{\TheWordProof\enskip}%
}%

\def\@endProof{%
   \qed\restore@qed
   \penalty-700 \smallskip
}

\def\@qed#1{\check@pt@fm@thm
\if@qed                                 % have we already done \qed?
     \global\@qedfalse\pushright{#1}%  - no  - do it now, but not again
\else\ifhmode\ifinner\else\par\fi\fi%  - yes - just end paragraph, if any
\fi}

\def\@pushright#1{%
  \ifvmode
       \ifinner\vbox@pushright{#1}% vertical mode (see comments above)
       \else   \page@pushright{#1}%
       \fi
  \else\ifmmode\maths@pushright{\hbox{#1}}% maths (force a text argument)
       \else\ifinner\hbox@pushright{#1}%  inside an \hbox
            \else\parag@pushright{#1}%    in a paragraph
  \fi  \fi  \fi
}% (22 Feb 1993) removed extra {} which would destroy our \] below

\def\maths@pushright#1{%
  \ifinner
     \hbox@pushright{#1}%
  \else
     \eqno#1%   use TeX's right equation number feature (\eqno) within $$.
     \def\]{$$\ignorespaces}%    suppress LaTeX's error checking (HACK!)
  \fi
}% (22 Feb 1993) removed extra {} which would destroy our \] above

\def\page@pushright#1{% 18 Jan 1994
  \skip@\lastskip
  \ifdim\skip@>\z@
       \unskip    % remove \parskip, but only (15 Feb 1994) if it's positive
  \else\skip@\z@\relax
  \fi
  \dimen@\baselineskip
  \advance\dimen@-\prevdepth            % save \prevdepth to make a strut
  \nobreak                              % don't break the page here
  \nointerlineskip
  \hbox to\hsize{%
    \setbox\z@\null
    \ifdim\dimen@>\z@\ht\z@\dimen@\fi   % simulate \baselineskip
    \box\z@
    \hfill
    #1}%
  \vskip\skip@                  % replace old \parskip
}%

\ifx\nonqed@thm\undefined % otherwise double loading causes a loop 22 Feb 1993
   \let\nonqed@thm\@thm % save original
   \let\nonqed@endthm\@endtheorem
\fi

\def\@thm{\save@set@qed\nonqed@thm}%
\def\@endtheorem{\ifAutoQED\qed\fi\restore@qed\nonqed@endthm}%

\newbox\qed@box % the box in which to save \qed for special handling

\def\WillHandleQED{\relax
\ifx\HandleQED\nohandle@qed
   \def\pushright{\global\setbox\qed@box\hbox}% \qed will be saved
   \let\QEDbox\qed@box % point at the right box
   \def\HandleQED{\unhbox\QEDbox}% print it
   \aftergroup\check@handle@qed  % check that this really gets done
\else
   \let\QEDbox\voidb@x % by pointing at an empty box
\fi}

\def\nohandle@qed{%
\errhelp{One of them is missing: see QED.sty.}%
\errmessage{This environment uses \string\WillHandleQED\space and
\string\HandleQED\space incorrectly}}

\def\check@handle@qed{\relax
\ifvoid\qed@box\else\expandafter\nohandle@qed\fi}

\def\UnHandleQED{%
\let\HandleQED\nohandle@qed
\let\QEDbox\voidb@x
\def\pushright{\protect\@pushright}}%

\UnHandleQED

\def\obsolete@pt@version#1{\errhelp={%
        Anonymous FTP ftp.dcs.qmw.ac.uk /pub/tex/contrib/pt/tex/#1}%*
\errmessage{You have an obsolete version of #1 - please get a new one}}%

\ifx\theorem@style\undefined
   \def\check@pt@fm@thm{\relax
     \ifx\square\undefined
       \gdef\square{\bigcirc
          \errhelp={Anonymous ftp e-math.ams.com /ams/amsfonts}%
          \errmessage{\string\square\space is an AMS symbol}%
          \global\let\square\bigcirc}%
     \fi
      \ifx\theorem@style\undefined
          \global\let\check@pt@fm@thm\relax
      \else\errhelp={The macros \@thm and \@endtheorem need to be re-defined.}%
          \errmessage{QED.sty must be loaded AFTER theorem.sty but before
           using \string\newtheorem}%
      \fi
      \global\let\check@pt@fm@thm\relax
      }%
\else
   \def\check@pt@fm@thm{%
      \ifx\square\undefined
         \def\square{\bigcirc
         \errhelp={Anonymous ftp e-math.ams.com /ams/amsfonts}%
         \errmessage{\string\square\space is an AMS symbol}%
         \global\let\square\bigcirc}%
      \fi
      \global\let\check@pt@fm@thm\relax
      }%
\fi

\def\check@pt@sty@v{%
   \relax
   \ifx\@@begintheorem\undefined
   \else\message{*** QED.sty and the old Paul.sty seriously conflict! ***}%
        \obsolete@pt@version{Paul.sty}%
   \fi
   \ifx\execHorizontalMap\undefined
        \gdef\obsolete@pt@version##1{}% it's no longer needed
   \else\global\let\di@gr@m\diagram
        \gdef\diagram{\obsolete@pt@version{diagrams.tex}\di@gr@m}%
   \fi
   \global\let\check@pt@sty@v\relax
}%

\def\DefineStandardTheorems{%
\relax\ifx\@@thm\undefined
        \@addtoreset{Result}{section}%
        \ifx\chapter\undefined\else\@addtoreset{Result}{chapter}\fi
\fi
\let\theUnnumbered\relax\countdef\c@Unnumbered255 \def\p@Unnumbered{}%
\let\DefineStandardTheorems\relax
}

\ifx\newtheorem\undefined\else\ifx
\else
\fi\fi

\ifx\theorem@style\undefined\else\qedrestoreat\expandafter \fi
\def\@opargbegintheorem#1#2#3{\begin@theorem{\TheoremHeader{#1}{#2}{ (#3)} }}

\def\@begintheorem#1#2{\begin@theorem{\TheoremHeader{#1}{#2}{} }}

\def\begin@theorem#1{%
  \trivlist\item[\hskip\labelsep {#1}]%         % LaTeX likes this method.
  \TheoremFont                                  % Italicise the enunciation
  \ifx\ProofFont\empty\def\ProofFont{\rm}\fi    % but not any nested proof.
  }%

\ifx\newtheorem\undefined\else\qedrestoreat\expandafter \fi
\def\ther@sult#1{%
\ifx\chapternumber\undefined\else
   \ifnum\chapternumber>0 \number\chapternumber.\fi\fi
\ifx\sectionnumber\undefined\else
   \ifnum\sectionnumber>0 \number\sectionnumber.\fi\fi
\ifx\subsectionnumber\undefined\else
   \ifnum\subsectionnumber>0 \number\subsectionnumber.\fi\fi
\expandafter\number\csname c@#1\endcsname}%

\def\plain@thm#1#2{% #1=counter, #2=text
  \save@set@qed
  \expandafter\advance\csname c@#1\endcsname1
  \@begintheorem{#2}{\csname the#1\endcsname}%
}%

\def\newtheorem#1[#2]#3{%   #1=command, #2=counter, #3=text
  \expandafter\def\csname #1\endcsname{\plain@thm{#2}{#3}}%
  \expandafter\def\csname end#1\endcsname{\@endtheorem}%
  \expandafter\ifx\csname c@#2\endcsname\relax
     \expandafter\expandafter\csname newcount\endcsname
        \csname c@#2\endcsname
     \expandafter\def\csname the#2\endcsname{\ther@sult{#2}}%
  \fi
}%

\TheoremsAsCommands

\qedrestoreat 

\newtheorem{observation}{\textbf{Observation}}

\newtheorem{definition}{\textbf{Definition}}
\newtheorem{assumption}{\textbf{Assumption}}
\newcommand{\ignore}[1]{}
\newif\ifspace\spacefalse
\newif\ifspacetwo\spacetwofalse

\usepackage{algpseudocode}

\usepackage{url}
\makeatletter
\g@addto@macro{\UrlBreaks}{\UrlOrds}
\makeatother

\newcommand{\hpt}{\textsf{hpt}\xspace}
\newcommand{\lpt}{\textsf{lpt}\xspace}
\newcommand{\cleanpt}{\textsf{clpt}\xspace}
\newcommand{\ubits}{\textsf{u}\xspace}

\newcommand{\newclock}{\textsf{pwc}\xspace}

\newcommand{\NEWCLOCK}{\textsf{PWC}\xspace}
\newcommand{\lastccc}{r\xspace}
\newcommand{\ccc}{\ensuremath{e_1, e_2, \cdots, e_\lastccc}\xspace}

\newcommand{\minsend}{\delta_{se}}
\newcommand{\minreceive}{\delta_{re}}
\newcommand{\mintransmit}{\delta_{tr}}
\newcommand{\minlocal}{\delta_{loc}}

\newcommand{\avtransmit}{av_{tr}}
\newcommand{\avcomp}{av_{comp}}

\newcommand{\QEDB}{\null\nobreak\hfill\ensuremath{\square}}%

\newcommand{\snote}[1]{{\color{purple}{TODO #1}}}

\newcommand{\dnote}[1]{{\color{blue}{#1}}}
\newcommand{\gnote}[1]{{\color{red}{#1}}}

\usepackage[ruled,vlined]{algorithm2e}

\newcommand{\correct}{viable\xspace}
\newcommand{\correctness}{viability\xspace}

\newcommand{\pwc}{\ensuremath{pwc}\xspace}

\newcommand{\br}[1]{\ensuremath{\langle #1 \rangle}\xspace}

\newcommand{\hb}{\ensuremath{\longrightarrow}\xspace}
\newcommand{\chb}{\ensuremath{\|}}
\newcounter{assumptions}
\newcounter{definitions}
\newcounter{implications}
\newcounter{lemmas}
\newcounter{observations}
\newcounter{theorems}

\begin{document}

%
% The "title" command has an optional parameter, allowing the author to define a "short title" to be used in page headers.
\title{Achieving Causality with Physical Clocks}

%
% The "author" command and its associated commands are used to define the authors and their affiliations.
% Of note is the shared affiliation of the first two authors, and the "authornote" and "authornotemark" commands
% used to denote shared contribution to the research.
\author{Sandeep S. Kulkarni}
\email{sandeep@msu.edu}
\affiliation{%
  \institution{Michigan State University}
  \city{East Lansing}
  \state{Michigan}
}

\author{Gabe Appleton}
\email{applet14@msu.edu}
\affiliation{%
  \institution{Michigan State University}
  \city{East Lansing}
  \state{Michigan}
}

\author{Duong Nguyen}
\email{nguye476@msu.edu}
\affiliation{%
  \institution{Michigan State University}
  \city{East Lansing}
  \state{Michigan}
}

%
% By default, the full list of authors will be used in the page headers. Often, this list is too long, and will overlap
% other information printed in the page headers. This command allows the author to define a more concise list
% of authors' names for this purpose.
\renewcommand{\shortauthors}{Kulkarni, Appleton, and Nguyen}

%
% The abstract is a short summary of the work to be presented in the article. 

\begin{abstract}
Physical clocks provide more precision than applications can use. For example, a 64 bit NTP clock allows a precision of 233 picoseconds.
In this paper, we focus on whether the least significant bits that are not useful to the applications could be used to track (one way) causality among events. 

We present \NEWCLOCK (Physical clock With Causality) that uses the extraneous bits in the physical clock. 
We show that \NEWCLOCK is very robust to errors in clock skew and transient errors.
We show that \NEWCLOCK can be used as both a physical and logical clock for a typical distributed application even if just 6-9 extraneous bits (corresponding to precision of 15-120 nanoseconds) are available.
Another important characteristic of \NEWCLOCK is that the standard integer $<$ operation can be used to compare timestamps to deduce (one-way) causality among events. 
Thus, \NEWCLOCK is significantly more versatile than previous approaches for using the physical clock to provide causality information. 
%\snote{This needs to be updated}
\end{abstract}



%
% Keywords. The author(s) should pick words that accurately describe the work being
% presented. Separate the keywords with commas.

%
% A "teaser" image appears between the author and affiliation information and the body 
% of the document, and typically spans the page. 
%%  \includegraphics[width=\textwidth]{sampleteaser}
%%  \caption{Seattle Mariners at Spring Training, 2010.}
%%  \Description{Enjoying the baseball game from the third-base seats. Ichiro Suzuki preparing to bat.}

%
% This command processes the author and affiliation and title information and builds
% the first part of the formatted document.
\maketitle

%\input{ieeestyle}
% \input{acmtext}

\iffalse
\begin{abstract}
Physical clocks provide more precision than applications can use. For example, a 64 bit NTP clock allows a precision of 233 picoseconds.
In this paper, we focus on whether the least significant bits that are not useful to the applications could be used to track (one way) causality among events. 

We present \NEWCLOCK (Physical clock With Causality) that uses the extraneous bits in the physical clock. 
We show that \NEWCLOCK is very robust to errors in clock skew and transient errors.
We show that \NEWCLOCK can be used as both a physical and logical clock for a typical distributed application even if just 6-9 extraneous bits (corresponding to precision of 15-120 nanoseconds) are available.
%
Another important characteristic of \NEWCLOCK is that the standard integer $<$ operation can be used to compare timestamps to deduce (one-way) causality among events. 
%
Thus, \NEWCLOCK is significantly more versatile than previous approaches for using the physical clock to provide causality information. 
%\snote{This needs to be updated}
\end{abstract}
\fi

%\gnote{Not sure how to do the CCS part, or why it is referencing ACM in the ``ACM Reference Format''}

\section{Introduction}

Computer physical clocks often provide precision that is more than what applications can use. For example, a typical clock used by NTP is 64 bits long~\cite{RFC0958NTP}. Of these, the last 32 bits represent a fractional second. In turn, it gives a theoretical resolution of $2^{-32}$ seconds (233 picoseconds). NTPv4 introduces a 128-bit clock of which 64 bits are for the fractional second. Thus, the smallest resolution is $2^{-64}$ seconds (0.05 attoseconds). %\todo{Similar statement}

%the last 16 bits provide precision of nano-seconds or below. \snote{Duong: Can you add details for Linux and something else} \dnote{I am looking at the time\_t data structure of C but it seems the granularity is second. I am chechking for more details.}

For any practical program in a distributed system, these lower-end bits are essentially useless, as they provide a precision that is well below what the application can use. For example, if the execution of an instruction takes $1 ns$, any time with finer granularity is meaningless. In fact, for many applications, even larger granularity is sufficient. As an illustration, consider a database application that needs to store the update timestamp of data. In this case, any clock bits that have finer granularity than the time to update an entry are redundant. Or in a monitoring program, if the monitor samples the state of a target application every 100 milliseconds, then granularity finer than $100ms$ is redundant to the monitor. Moreover, if an application is written in Java or Python, then the smallest clock resolution that it could query the system for is 1 nanosecond, and the bits corresponding to values less than $1ns$ are not used.
%\gnote{Editing: We have two sentences nearly in a row starting with ``for example"}
%
%\gnote{Should we use "clock resolution" instead of "time accuracy"? Duong: done}
%
Thus, if we zero out any extraneous bits every time the application uses them, the application would not be affected.
%\snote{see now. D: I'm good}
%\snote{Duong: Can you add one or two examples on this. Yes, I have added}

A natural question is whether we can utilize these bits to achieve something useful. In particular, our focus is on using these bits so the physical clock can be simultaneously used as a logical clock \cite{Lamport1974CACM}.
%building counters that provide causal information similar to that provided by logical clocks \cite{Lamport78CACM}. 
In this case, the information would be useful to clients that are aware of the purpose of these bits, and clients unaware of the purpose of these bits would not be affected. 
%\todo{S:Rem}
%This is due to the fact these applications could not distinguish between these bits with the granularity under consideration. 

\subsection{Why Incorporate Causality in Physical Clocks}

Distributed systems lack a common global clock and processes can only use a local (physical) clock. Since these clocks may grow at their own rate, it presents several challenges that have been identified in the literature. 

For example, in Spanner \cite{CDEFFFGGHH13TOCS}, the authors use tightly synchronized physical clocks to ensure that if we have two transactions $T1$ and $T2$ then the start time of $T2$ is strictly ahead of that of $T1$ even if $T1$ and $T2$ are running on two different machines. This introduces a commit-wait delay, which occurs when $T1$ is writing data that is read by $T2$, but the clock of the machine running transaction $T2$ is lagging.  

%As another example, GentleRain \cite{GentleRain.DIRZ14SOCC} is used to provide causal consistency using partially synchronized physical clocks that require that if update of variable $x$ happens before the update of variable $y$ then the timestamp of $y$ must be higher than that of $x$ even if $x$ and $y$ are on two different machines. \gnote{Not sure how to immediately fix, but this is a run-on sentence. Needs breaking up somehow.} To achieve this, write operation of $y$ may be delayed until the clock of the machine hosting $y$ catches up. 

As another example, GentleRain \cite{GentleRain.DIRZ14SOCC} provides causal consistency using 1) partially synchronized physical clocks, and 2) requiring that if the update of variable $x$ happens before the update of variable $y$, then the timestamp of $y$ must be higher than that of $x$ even if $x$ and $y$ are on different machines. To achieve this, the write operation of $y$ needs to be delayed until the clock of the machine hosting $y$ catches up. 
%\gnote{Changed sentence a little bit. Is this still accurate? Original is commented above.}
%to satisfy this constraint. 

%If physical clocks were extended to support causal order then these problems would go away.
If it were guaranteed that physical clock of $f$ was higher than that of $e$ whenever event $f$ is causally dependent on event $e$, these problems would go away. 
For example, in the Spanner example above, the timestamp of $T2$ would be automatically higher than that of $T1$ thereby obviating the need for a commit-wait delay. 
%For example, in case of Spanner, if it was guaranteed that time of $T2$ would be higher than that of $T1$ there would be no commit-wait due to the causality between $T1$ and $T2$. 
CockroachDB has used this approach to eliminate such delays.
%demonstrated that incorporating physical clocks to have causality, it is possible to eliminate such delays. 
Likewise, in the case of the GentleRain example, the timestamp of update of $y$ would be automatically higher than that of $x$, thereby eliminating latency in write operations. 
%Likewise, in GentleRain if the clock of $y$ was guaranteed to be ahead of that of $x$, there would no increased latency for write operations. 
CausalSpartan  \cite{CausalSpartan.RDK17SRDS} demonstrated this where they showed that elimination of the write-delay reduces response times to certain queries by an order of magnitude.
%even a clock skew of $2ms$ can increase response time of some of the queries five-fold. %\sandeeptodo{Look at the multiplier effect to say something}

Providing physical clocks the ability with to capture causality between events also provides additional avenues to utilize them. For example, using logical clocks, it is possible to take a snapshot of a system in a trivial manner; if each process takes a snapshot when its (logical) clock value is $T$ (a pre-defined value) then the resulting snapshot is consistent. However, the same is not true for (ordinary) physical clock. This ability to take snapshot with logical clocks is not useful in practice as different processes may reach logical clock value $T$ at very different times. On the other hand, an ability to take a snapshot at physical time $T$ would be valuable; it could be used to take a snapshot in the past (e.g., snapshot of the system at 5pm the day before) or in the future (every day at 5pm). This idea is used in \cite{CADK17ICDCS} so that we can take snapshots at will, rollback the system to its state a few minutes ago, etc. It has been demonstrated that Retroscope can take up to 150,000 snapshots per second and can be used to monitor a distributed key-value store and perform on-demand rollback without affecting other clients. 
%\sandeeptodo{One/two more sentences about what the paper does}

A key requirement to allow physical clocks to detect causality is backward compatibility, i.e., the structure of the physical clock (a 64/128 bit integer) should not change. A more complex structure would make it difficult to use them in practice as the clock value is part of several data structures and changing them is not feasible or advisable.

\subsection{Prior Approaches}

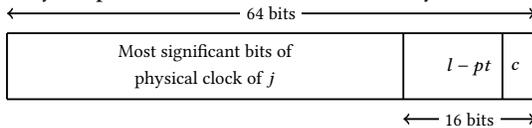
\begin{figure}[h]
\vspace*{-3mm}
\small
\centering
\resizebox{0.4\textwidth}{!}
%\resizebox{5cm}{1cm}
{
\begin{tikzpicture}
\draw[thick] (0,1) -- (8,1);
\draw[thick] (0,0) -- (8,0);
\draw[thick] (0,0) -- (0,1);
\draw[thick] (6,0) -- (6,1);
\draw[thick] (8,0) -- (8,1);
\draw[thick, <->] (0,1.3) -- (8,1.3);
\draw (4,1.3) node [fill=white] {64 bits};
\draw[thick, <->] (6,-0.3) -- (8,-0.3);
\draw (7,-0.3) node [fill=white] {16 bits};

\draw (3,0.7) node {Most significant bits of};
\draw (3,0.3) node {physical clock of $j$};
\draw (7,0.5) node {$l-pt$};
\draw[thick] (7.5,0) -- (7.5,1);

\draw (7.7,0.5) node {$c$};
\end{tikzpicture}
}
\vspace{-10pt}
\caption{Format of 64-bit Hybrid Logical Clock of an event at process $j$}
\label{fig:hlc-format}
\vspace*{-16pt}
\end{figure}

Hybrid logical clocks (HLC) \cite{KDMA2014OPODIS.HLC} provide one such approach to enable physical clocks to provide causal information. In \cite{KDMA2014OPODIS.HLC}, the clock consists of three parts, $pt$, $l$, and $c$, where $pt$ is the physical clock where the event occurred,  $l$ is close to the physical clock, and $c$ is a counter whose value is generally very small. In \cite{KDMA2014OPODIS.HLC} it is proposed that the last 16 bits of $pt$ 
%\dnote{(the last 16 bits of HLC?)}
%\snote{no. pt is correct}
are used to store the value of $l-pt$ and the value of $c$ (cf. Figure~\ref{fig:hlc-format}). With this approach, a client who is aware of this encoding can decode to obtain values of $l$ and $c$ that are needed to compare causality between events. 
While we discuss the details of HLC in the next section, we note that one disadvantage of this approach is that applications that use these clocks need extra work to identify the precise value of $l$ and $c$ before they can be used.

%\snote{I will use this for marking text that I am removing for space. Check here to see if you are ok with it. Add your name to it as in S: G: D: Removed. Then, we can decide to add some of it back if we have space. }

% \begin{figure}
% \centering
% \begin{tikzpicture}
% \draw[thick] (0,1) -- (8,1);
% \draw[thick] (0,0) -- (8,0);
% \draw[thick] (0,0) -- (0,1);
% \draw[thick] (6,0) -- (6,1);
% \draw[thick] (8,0) -- (8,1);
% \draw[thick, <->] (0,1.3) -- (8,1.3);
% \draw (4,1.3) node [fill=white] {64 bits};
% \draw[thick, <->] (6,-0.3) -- (8,-0.3);
% \draw (7,-0.3) node [fill=white] {$u$ bits};

% \draw (3,0.5) node {Pseudo-Physical Time};

% \draw (6.7,0.5) node {$c$};
% \end{tikzpicture}
% \caption{Format of 64-bit Pseudo-Physical Clock}
% \label{fig:pwc-format}
% \end{figure}

\subsection{Our Approach}
Our focus in this paper is to develop a new type of clock, denoted as \NEWCLOCK, {physical clock with causality}. The goal of \NEWCLOCK is to begin with a parameter $\ubits$ that identifies the extraneous/useless/redundant 
%\dnote{(do we keep three words?)} 
%\snote{I like to keep it to say that they are synonymous}
bits and develop a clock that can be simultaneously used as a physical clock as well as a logical clock. 
Furthermore, we want to obviate the need for encoding and decoding when comparing timestamps, i.e., we should simply use \textit{integer comparison} to compare two clock values to identify which value is larger. No decoding/encoding should be necessary beyond applying bitmasks.

\textbf{Contributions of the paper. } 
We begin in a system where physical clocks of processes are within $\epsilon$ of each other, where $\epsilon$ is a given parameter. This parameter is not known to processes but is only required for ensuring correctness. We develop \NEWCLOCK, using a second parameter $\ubits$, which identifies the number of extraneous least-significant bits that are available. \NEWCLOCK has the following features. 
%\vspace*{-4mm}
\begin{itemize}
    \item It can be used as a substitute for the physical clock. Specifically, for any process $j$, the value of $\newclock.j$ is \textit{close} to the physical clock of $j$. 
    \item It can be used as a substitute for a logical clock. Specifically, for any two events $e$ and $f$, if event $e$ causally affects $f$ (cf. Section \ref{sec:prel}) then $\newclock.e < \newclock.f$, where $<$ is the standard ``less than'' relation between integers.
    %\item To determine $\newclock.e < \newclock.f$, only an integer comparison is sufficient
    \item \NEWCLOCK can be customized based on the number of useless/
    {extraneous} bits that are available. \NEWCLOCK can be adapted to situations where sufficient extraneous bits are unavailable or the number of extraneous bits needs to be changed dynamically. It is resilient to errors in the clock synchronization protocol.
   % \item If the number of available bits is insufficient, \NEWCLOCK can be adapted for those conditions.
   \item
   We analyze \NEWCLOCK and find that \NEWCLOCK can be used as physical clock and logical clock even in applications that require a very fine clock resolution (tens of nanoseconds). This implies that \NEWCLOCK can be used in virtually any application. By contrast, HLC \cite{KDMA2014OPODIS.HLC} cannot be used in applications that need such fine-grained clock resolution. Thus, \NEWCLOCK is more versatile than HLC. 
   
   %We analyze \NEWCLOCK and find that \NEWCLOCK can be used even in applications that need clock resolution of $10\!-\!100ns$. This implies that \NEWCLOCK can be used in virtually any application. 
%   \snote{I have kept 100 ns because we needed 9 bits in some place} \dnote{(Yes, I understand that the more $\ubits$ bits the larger resolution value supported by \NEWCLOCK. But I think the sentence is about resolution needed by application, not about resolution supported by \NEWCLOCK. With 9 bits, \NEWCLOCK can still support application which requires more than 1 ms resolution.)} \gnote{I read that sentence the same way Duong does. I think we should say $\ge 10$ unless there is another reason not to.}
%\item We find that for a typical distributed application, 5 extraneous bits are sufficient to satisfy the constraints of \NEWCLOCK. 
    %In the context of NTP, these 5 bits correspond to granularity of 7.5 nanoseconds. 
    %In the context of NTP, without these 5 bits, the smallest clock resolution supported is 7.5 nanoseconds.
    %Hence, \NEWCLOCK can be used in virtually any application. \snote{Need to change: let's talk}
    
\end{itemize}

\textbf{Organization of the paper. }
The rest of the paper is organized as follows: In Section \ref{sec:prel}, we define distributed systems, clock synchronization, and causality. We also identify the assumptions made in the paper. In Section \ref{sec:algorithm}, we define \NEWCLOCK and identify conditions under which the available extraneous bits are sufficient so that \NEWCLOCK can be used as a substitute for physical clock and logical clock. In Section \ref{sec:reliable}, we discuss the reliability of \NEWCLOCK. In Section \ref{sec:experiments}, we implement a simulated network environment and test the behavior of \NEWCLOCK, followed by a similar experiment on a physical network. We discuss several questions associated with \NEWCLOCK in Section \ref{sec:discussion}. We discuss related work in Section \ref{sec:related-work} and conclude in Section \ref{sec:concl}. For reasons of space, proofs, and additional discussions are in the Appendix.
%\snote{remaining sections}
% \dnote{(Good now. Check for final version)}

\section{Preliminaries}
\label{sec:prel}

\subsection{Distributed Systems}

We consider a distributed system that consists of $N$ processes communicating via messages. The state of a process is changed by (send/receive or local) events.
%To help with the step-by-step reconstruction of a distributed computation%(e.g. determine the ordering between events, especially ordering of events on different processes, and the sequence of state changes)
Each event is associated with a timestamp. Ideally, the timestamp of an event is the time the event occurs in \textit{absolute (global) time}. 
% \todo{S: Rem. D: Yes}
%Such a perfect timestamp will allow us to determine the order between any two events. 
However, absolute time is not practical since it requires infinitesimal granularity and it is not possible to have perfect clock synchronization among processes.

In practice, the timestamp associated with an event comes from the reading of the local clock of the machine/processor/process on which the event occurs.
%In practice, an event is timestamped from the reading of the local clock of the process on which the event occurs.
Processes use a protocol such as Network time protocol (NTP)~\cite{RFC0958NTP, RFC5905NTPv4} or Precision Time Protocol (PTP)~\cite{ieee1588-2002PTP}
to keep their local clock synchronized to a master clock and avoid the possibility of arbitrary clock skew due to hardware errors and other failures.
%
%Each process $j$ ($1 \le j \le n$) keeps track of the time by having its own local clock. The processes use a protocol such as Network Time Protocol (NTP \cite{RFC0958NTP, RFC5905NTPv4}) to synchronize their clocks. 
%We assume that the maximum difference between the local clocks of any pair of processes is $\epsilon$. 
%other Clock synchronization protocols like NTP ensure that the clock of the process is \textit{close} to the master clock as well as to the clocks of other processes.
We abstract this property by the requirement that clocks of two processes differ by at most $\epsilon$.
{The value of $\epsilon$ is anywhere between sub-microsecond to several milliseconds based on network size and connectivity. 
}
%\todo{S: Rem. D: Yes G: yes}
\begin{comment}

In the case of NTP, the clock synchronization error is typically within tens of microseconds to tens of milliseconds based on the network size, underlying network connectivity, and the stability of network nodes/links.
%tens of microsecond and within a few tens of millisecond in the case where network condition is unstable \todo{Duong: to check} (however, when network condition recovers to normal, the synchronization error is improved). 
Other protocols such as Precision Time Protocol could provide tight clock synchronization at sub-microsecond levels \cite{ieee1588-2002PTP}. Based on the above observation and the assumption that network failures are transient, we make the following assumption about the clock synchronization error:
\end{comment}
Hence, we make the following assumption: 

%\snote{This should be assumption. Counters are not correct}\dnote{fixed}

%\refstepcounter{assumptions}\textbf{Assumption \theobservations}. 
\refstepcounter{assumptions}\textbf{Assumption \theassumptions}. 
\label{asm:epsilonexist}
 There exists a bound $\epsilon$ such that the physical clocks of two processes differ by at most $\epsilon$.

We note that the value of $\epsilon$ is not required to be known to processes. The \NEWCLOCK algorithm does not use it in the implementation; it is needed only for the correctness requirement of \NEWCLOCK algorithm. 
% \snote{Changed}

%Initially, we focus on systems where the value of $\epsilon$ is known to the processes. In Section \ref{sec:discussion}, we consider the case where the value of $\epsilon$ cannot be determined or the clock synchronization assumption is violated. 

% \todo{S,D:Rem}
%\snote{TODO: Remove? There are three types of events that could occur on a process: local events, send events, and receive events. Send and receive events correspond to the communication between processes. Local events are internal events of processes in which they (may) change their local states. 
%}\dnote{(If we remove, it might affect the clarity of Algorithm~\ref{alg:timestamping} but I am OK with removal)}

\begin{comment}

\refstepcounter{assumptions}\textbf{Assumption \theassumptions}.
\label{asm:atmostonelocaleevent}
There is at most one \textit{local} event between two \textit{non-local} (send/receive) events.

The assumption \ref{asm:atmostonelocaleevent} is reasonable in that suppose there are multiple events $e_1$ and $e_2$ on process $j$ and $j$ neither sends nor receives a message between these two events then no other process knows the order between them. We can, therefore, assign $e_1$ and $e_2$ the same timestamp. \todo{check if needed}

\end{comment}
%\snote{I have removed assumption about at most one local event. Let's see if it is needed}

%Between the moment a process $j$ sends a message $m$ and the moment a process $k$ receives that message $m$,
The transmission of a message $m$ from $j$ to $k$ can be divided into three time intervals: the message $m$ is sent by originating process $j$ to its network interface card, $m$ is transmitted over the communication network, and $m$ is delivered to target process $k$ from its corresponding network interface card. Those time intervals are denoted as send time, transmit time, and receive time, respectively. We make the following assumption about those time intervals:

\refstepcounter{assumptions}\textbf{Assumption \theassumptions}.
\label{asm:deltasend}
Let $\minsend, \minreceive, \mintransmit,$ and $\minlocal$ denote the minimum time required to send a message, receive a message, transmit a message, and complete the tasks in a local event. We assume that each of these terms is greater than $0$. 

\begin{comment}

\begin{itemize}
    \item 
$\minsend, \minreceive, \mintransmit,$ and $\minlocal$ denote the minimum time required to send a message, receive a message, transmit a message, and create a local event. We assume that each of these terms is greater than $0$. 

The time required to send a message/receive (in absolute time) is at least $\minsend$ and $\minsend > 0$.
\item The time required to transmit a message (in absolute time) is at least $\mintransmit$ and $\mintransmit > 0$.
\item The time required to receive a message (in absolute time) is at least $\minreceive$ and $\minreceive > 0$.
\item The time between two consecutive events on a process is at least $\minlocal$ and $\minlocal > 0$.

\end{itemize}
\end{comment}

%Note that we assume processes reside on different machines (otherwise they could communicate via shared memory), and thus $\mintransmit > 0$. If multiple processes are running the same machine then we can use the physical clock of that machine for timestamping. If two events $e_1$ and $e_2$ are on two different processes running on the same machine and $e_1$ happened before $e_2$ then the physical clock of that machine would be sufficient to detect the order between those events. 
%\todo{Professor: do we need to assume processes are not on the same machines, otherwise $\mintransmit$ could be 0?}
%\sandeeptodo{I am not a fan of this text. Let's see if we can remove it. }

\subsection{Causality}

%Happened before and concurrent relationship, and their properties.

% \todo{S: Rem. D: Yes}
% \snote{Duong: You misunderstood. I am ok with hb but when I tried to change it, it did not work.}
%If local clocks of processes are perfectly synchronized to the absolute time, then the absolute timestamp can determine the ordering of events in a distributed system. However, such perfect synchronization is not possible in practical distributed systems. Thus, to determine a partial ordering among the events, we rely on the notion of causality \cite{Lamport78CACM} as defined below:

Let $e$ and $f$ be two events. We say that event $e$ happens before event $f$, denoted as $e$ \hb $f$, if and only if:
%\begin{itemize}
 %   \item
    (1) $e$ and $f$ are events on the same process and $e$ occurred before $f$, or
  (2) $e$ is a send event and $f$ is the corresponding receive event, or
    (3) There exists an event $g$ such that $e \hb g$ and $g \hb f$.
%\end{itemize}
%
%The notion $e \hb f$ indicates that there is a \textit{potential} causal relationship between $e$ and $f$. When no causal relationship exists, i.e. neither $e \hb f$ nor $f \hb e$ is true, 
We say $e$ and $f$ are concurrent, denoted as $e \chb f$, where $e \chb f \iff \neg (e \hb f) \wedge \neg (f \hb e)$.

\begin{figure}[h]
\centering
\vspace{-5pt}
\begin{subfigure}[b]{\linewidth}
    \centering
    \includegraphics[width=0.8\textwidth]{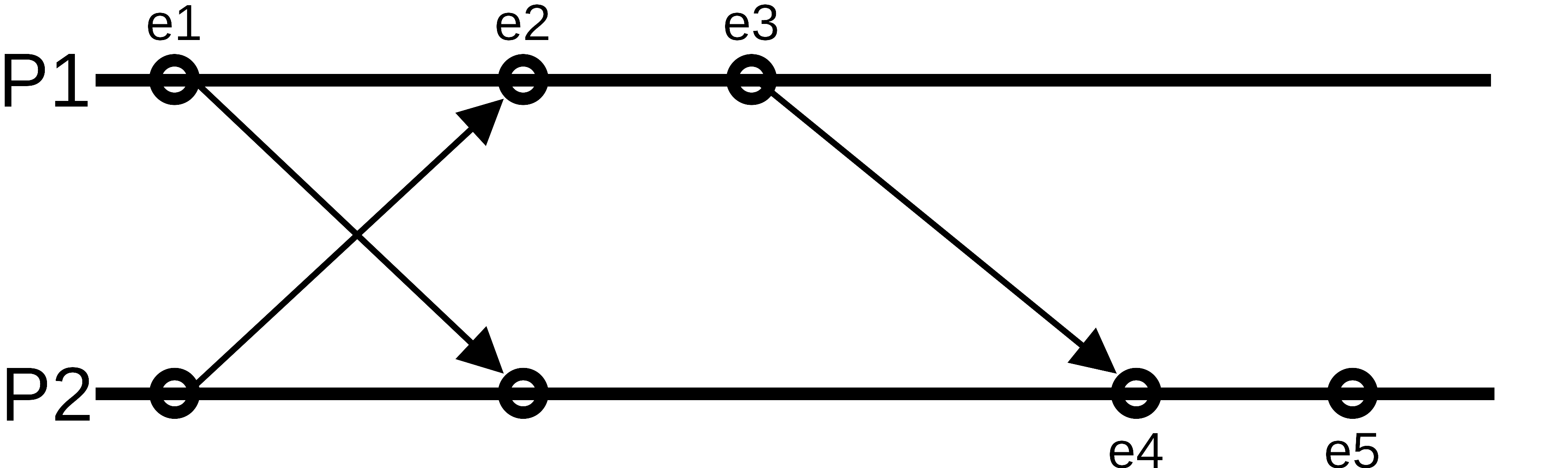}
    \subcaption{Example of Abstract Computation}
    \label{fig:abstractcomputation}
\end{subfigure}
\hfill
\begin{subfigure}[b]{\linewidth}
    \centering
    \vspace{1em}
    \includegraphics[width=\textwidth]{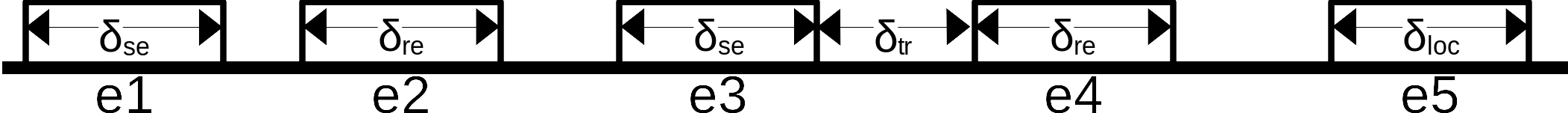}
    \subcaption{Example of Concrete Computation}
    \label{fig:concretecomputation}
\end{subfigure}
\caption{Example of Computation.}
\label{fig:computation}
\vspace{-14pt}
\end{figure}

A computation of a program is obtained by a partial order of the system's events such that the order respects causality. 
As an example, consider the computation shown in Figure \ref{fig:computation}. Figure \ref{fig:abstractcomputation} considers the traditional view of events in distributed computing where events are thought of as instantaneous. In practice, they take some non-zero time. For example, Figure \ref{fig:concretecomputation} shows the timeline of (a subset of) events from Figure \ref{fig:abstractcomputation}.% \snote{Gabe: Please fix this}

% Gabe thorough editing left off here
\subsection{HLC}

In this section we provide a brief overview of hybrid logical clocks \cite{KDMA2014OPODIS.HLC}. In hybrid logical clocks, each process $j$ maintains $pt.j$, which is the physical clock (system clock) of $j$, $l.j$ and $c.j$. 
The value of $pt.j$ is updated automatically by protocols such as NTP so that, at any given time, $|pt.j-pt.k|$ is bounded by clock skew $\epsilon$ for any two processes $j$ and $k$.

%The values of $l$ and $c$ are updated by the protocol given in Algorithm \ref{alg:HLC}.

While the detailed algorithm from \cite{KDMA2014OPODIS.HLC} is provided for completeness in Appendix, the key observations from \cite{KDMA2014OPODIS.HLC} are that (1) for any event $e$, $0 \leq l.e-pt.e \leq \epsilon$, and (2) the value of $c.e$ is bounded. In theory, the bound on $c$ is linear in $n$, the number of processes, and $\epsilon$. However, in practice, it is typically small. For this reason, in \cite{KDMA2014OPODIS.HLC}, it is proposed that the HLC timestamp can be saved as shown in Figure \ref{fig:hlc-format}. Here, in the 64-bit NTP timestamp, the last 16 bits are used to save the value of $l-pt$ (typically 12 bits) and $c$ (typically 4 bits).
%\snote{Sandeep: revisit}
%\snote{Gabe: add. If possible, add application that would have trouble. }

\section{\NEWCLOCK Algorithm}
\label{sec:algorithm}

In this section, we describe our algorithm to capture causality using the \textit{extraneous} bits, denoted by $\ubits$, in the physical clock. 
%As discussed in the introduction, we aim to utilize the \textit{extraneous} bits in the physical clock to provide causality. 
%Hence, one of the inputs that our algorithm uses is the number of available bits that are unneeded in the physical clock. 
This section is organized as follows: Section \ref{sec:newclockalgorithm} describes the algorithm. Since the algorithm requires 
\textit{sufficient extraneous bits} to be available, Section \ref{sec:correct} identifies sufficient conditions on the required number of extraneous bits. 
Subsequently, in Sections \ref{sec:logical} and \ref{sec:physical}, we show that \NEWCLOCK can be used as a substitute for logical clock and physical clock respectively. 

\subsection{\NEWCLOCK Algorithm}
\label{sec:newclockalgorithm}

We use the term \textit{\cleanpt.j} to denote the clock of a process or event $j$ where the extraneous $\ubits$ bits are masked to $0$. 
% \todo{S:Rem.D:yes}
%\snote{text to remove? As an illustration, if the physical clock is $1101011101$ (in binary) and it is determined by the user that 3 bits are useless/extraneous/redundant, the corresponding \textit{\cleanpt} is $1101011000$.
%}
{We assume that each process $j$ begins with the \textit{first} event, say $e$ such that $\newclock.j = \newclock.e = \cleanpt.e$.}
%the physical clock as well as the value of \newclock is $0$ for that event}

In our algorithm, each process $j$ maintains a variable $\newclock.j$ that keeps track of the timestamp of $j$ as well as the timestamp of the most recent event on $j$. To create a local event or to send a message, it increments the $\newclock$ value by $1$. It also takes a maximum of the $\newclock$ value with \textit{\cleanpt} value.  
%This timestamp is sent with any message sent by the process 
This timestamp is piggybacked on any message $m$ sent by the process, and is denoted as $\newclock.m$.
The receive operation also works similarly. $\newclock.j$ is set to be larger than the old value of $\newclock.j$, $\newclock.m$ and at least equal to $\cleanpt.j$ (the physical clock of $j$ with extraneous bits reset to $0$).
Thus, our algorithm is as follows:

\vspace{-10pt}
\begin{algorithm}[ht!]
\SetAlgoLined
\KwResult{Compute $\newclock$ value for newly created event $e$}
Initialization: $\newclock.j = \cleanpt.j$

\If{$e$ is a local event}
{
$\newclock.j = max(\newclock.j+1, \cleanpt.j)$\;
}
\If{$e$ corresponds to sending of message $m$}
{
$\newclock.j = max(\newclock.j+1, \cleanpt.j)$\;
$\newclock.m = \newclock.j$\;
}

\If{$e$ corresponds to receiving a message $m$ with timestamp $\newclock.m$}
{
$\newclock.j = max(\newclock.j+1, \newclock.m+1, \cleanpt.j)$\;
}
$\newclock.e = \newclock.j$\;
 \caption{Capturing causality with Extraneous Bits in Physical Clock}
 \label{alg:timestamping}
\end{algorithm}
\vspace{-10pt}

%\todo{use u.j for clock?}

% \todo{S: Rem.D:yes}
%A reader may observe that this algorithm is in fact similar to the naive algorithm in \cite{KDMA2014OPODIS.HLC} except that it uses \textit{\cleanpt.j} instead of the physical clock of $j$. While the naive HLC algorithm can suffer from scenarios the drift between $l$ and $pt$ values grows unbounded, in this paper, we focus on identifying  conditions that avoid the problem identified in \cite{KDMA2014OPODIS.HLC}.

%Note that since \cleanpt and \newclock are both derived from the NTP format, they inherit the NTP fractional second as a unit, each equal to $\tfrac{1}{2^{32}}$ seconds. If the implementing system uses NTPv4, then this unit may instead be $\tfrac{1}{2^{64}}$ seconds.

Note that \cleanpt and \newclock are represented in the same format as an NTP clock rather than as a floating point value as in some standard libraries. Conversion to other formats can be done in the same manner as done for NTP clocks.

\subsection{Sufficient Conditions for \correctness of Algorithm \ref{alg:timestamping}}
\label{sec:correct}

Our next goal is to identify under what conditions can Algorithm \ref{alg:timestamping} be \correct.
To define the notion of \correctness, we view the timestamp \newclock to consist of two parts: \hpt and \lpt. 
The least significant $\ubits$ bits are used for \lpt whereas the remaining bits are used for \hpt. 
% \todo{S:Rem.D:y}
%Specifically, \hpt corresponds to the \textit{most significant} bits of \newclock and $\lpt$ denotes the \textit{least significant} bits of \newclock. The size of \hpt and \lpt would depend upon the number of extraneous bits available in the physical clock. Specifically, if the physical clock consists of 64 bits and 12 least significant bits are thought of as unneeded based on the precision of clocks that an application could use then we use 12 bits for \lpt and 52 (i.e., 64-12) bits for \hpt. 

The key idea in defining \correctness is that the bits represented by \lpt are irrelevant as far as the physical clock is concerned.  $\NEWCLOCK$ permits $\lpt$ to be higher than $0$. As long as this value does not \textit{overflow}, we can use the proposed timestamp. Note that if this value overflows then it would affect \hpt, and it would impact the process as it would effectively change the physical clock being seen by the process.  With this intuition, we focus on defining the notion of \correctness, next. 
We begin with the following assumption.

\refstepcounter{assumptions}\textbf{Assumption \theassumptions}.
\label{asm:ubitspositive}
We assume that the number of extraneous bits is greater than $0$, i.e., $\ubits > 0$. 

\refstepcounter{assumptions}\textbf{Assumption \theassumptions}.
\label{asm:ubitsconstant}
We assume that the value of $\ubits$ is constant and identical across all processes during a computation. We consider the case where this value is changed dynamically in Section \ref{sec:dynamicu}. 

\refstepcounter{assumptions}\textbf{Assumption \theassumptions}.
\label{asm:monotonic}
In the initial discussion, we assume that physical clock of process $j$ is monotonic. We consider the issue of non-monotonic clocks in Section \ref{sec:nonmonotonic}. 

%The next assumption captures the notion of \textit{extraneous bits}. \snote{Appendix?}\dnote{(Assumption~\ref{asm:oneeventperclocktick} seems not directly mention \ubits)}

%Intuitively, we expect that clocks are incremented at least once between two events. Furthermore, based on the intuition of \textit{extraneous} bits, the significant bits of the physical clock must increase by at least one between events. 

\refstepcounter{assumptions}\textbf{Assumption \theassumptions}.
\label{asm:oneeventperclocktick}
The number of events created on process $j$ for a given $\hpt.j$ value is bounded, i.e., 
for any $t$, $| \{e | e$ is an event on $j$ and $\hpt.e = t\} |$ is bounded by a constant. For sake of simplicity, we assume that this bound is 1. 
{We note that if this bound is $K$ then $\lceil \log_2 K \rceil$ bits of $\lpt$ would be needed to distinguish such events thereby reducing the extraneous bits. }
%\dnote{(Is it because of that the clock resolution provided by $\hpt$ is small/good enough so that events will have distinct timestamp?)}. 
Thus, for \textbf{any two consecutive events} $e$ and $f$ on process $j$, we have 
$\hpt.e < \hpt.f$

We note that the impact of this assumption can be reduced by allowing broadcast messages where a node sends multiple messages simultaneously if all these events are timestamped with the same timestamp. Likewise, if a process receives multiple messages \textit{simultaneously}, then those receive messages can be combined into one receive event. In our implementation, we do not consider this optimization. Using such an optimization would improve applicability of \NEWCLOCK by reducing the number of extraneous bits required for correctness

Finally, the next assumption combines clock skew assumption \ref{asm:epsilonexist} and requirement of finite time spent in each event. Specifically,

\refstepcounter{assumptions}\textbf{Assumption \theassumptions}.
\label{asm:epsilonmeansseparate}
We assume that if absolute time $\epsilon$ has passed between events $e$ and $f$ then $\hpt.e  < \hpt.f$

%\dnote{In my understanding, we are assuming $\minlocal \ge {2^{\ubits}}$. Do we need to assume similar assumption for $\minsend, \mintransmit, \minreceive$ for the proof of Observation~\ref{obs:syncclocks}?}

%\snote{I am removing the assumption about no local events. It would need changes to other parts of the paper. }

\begin{comment}

\refstepcounter{assumptions}\textbf{Assumption \theassumptions}.
\label{asm:sendfollowedreceive}
For sake of simplicity, in the initial discussion, we assume that there are no local events. 
In other words, an event is either a send event or a receive event. 
%Furthermore, if a process has consecutive send events (respectively,  consecutive receive events) then those send events (respectively, receive events) are treated as one event. In other words, if the system execution is as shown in Figure \ref{??} then it is treated as that in Figure \ref{??}. Effectively, it will cause events \ref{??} to have the same timestamp. We note that this assumption makes the proof of the algorithm simpler. However, it can be removed as discussed in Section \ref{sec:discussion}.

\end{comment}

Based on the definition of \hpt and \cleanpt, $\cleanpt.e = 2^\ubits \cdot \hpt.e$. Thus, we have

\refstepcounter{observations}\textbf{Observation \theobservations}.
\label{obs:hptincrease}
Let $e$ and $f$ be two consecutive events on process $j$. Then, based on Assumptions \ref{asm:ubitspositive} and  \ref{asm:oneeventperclocktick}, we have 

$\cleanpt.e + 2^{\ubits} \leq \cleanpt.f$ 

$\cleanpt.e + 1 \leq \cleanpt.f$ %\dnote{(Indeed since $\ubits >0 \Rightarrow 2^{\ubits} > 1 \Rightarrow \cleanpt.e + 1 < \cleanpt.f $)} 

% \todo{S,D: Rem}
%If the physical clocks are perfectly synchronized then the least significant bits of \newclock will always be $0$. This is because the value of $\cleanpt$ would be highest in the max operation performed in the algorithm. Specifically,

\refstepcounter{observations}\textbf{Observation \theobservations}.
\label{obs:syncclocks}
If clocks were perfectly synchronized then for any event $e$, \lpt.e = 0

For reasons of space, we have kept proofs in Appendix.

If clocks are not perfectly synchronized then the value of $\lpt$ could be higher than $0$.
%\todo{S:Rem. D:yes G: yes}
%{\snote{Remove this text and figure? One way it could happen is as shown in Figure \ref{fig:lptgreaterthanone}. 
%}
In this case, we can make the following observation:
% \todo{S: Rem. D: yes}
\begin{comment}

\begin{figure}
\centering
\resizebox{4cm}{2cm}{
\begin{tikzpicture}
\draw[thick] (0,1) -- (4,1);
\draw[thick] (0,2) -- (4,2);
\draw[thick, ->] (0.5,1) -- (1,2);
\draw (1.5,0.1) node [fill=white] {\textit{Physical time:} 10101011-000};
\draw (1.5,0.6) node [fill=white] {\textit{\newclock:} 10101011-000 };

\draw (1.5,2.5) node [fill=white] {\textit{Physical time:} 10101010-000};
\draw (1.5,3.0) node [fill=white] {\textit{\newclock:} 10101011-001};

\end{tikzpicture}
}
\caption{Need for $\lpt > 0$. An hyphen is added between the values of \hpt and \lpt for ease of understanding.}
\label{fig:lptgreaterthanone}
\end{figure}
\end{comment}

\refstepcounter{lemmas}\textbf{Lemma \thelemmas}.
\label{lem:lptfnonzeo}
If there exists an event $f$ such that $\lpt.f > 0$ then there exists an event $e$ such that $e$ happened before $f$ and $\newclock.e = \newclock.f - 1$

We can generalize this observation recursively. Specifically, if $\lpt.f =5$ it implies that there is an event with timestamp $\newclock.f -1$. For this event, $\lpt$ value is 4. Thus, there is another event with timestamp $\newclock.f-2$ ($\lpt = 3$) and so on. Thus, we have

\refstepcounter{lemmas}\textbf{Lemma \thelemmas}.
\label{lem:chain}
If there exists an event $f$ such that $\lpt.f = v > 0$ then there exists events $e_1, e_2, \cdots e_{v}$, 

\begin{itemize}
    \item $\forall w : 1 \leq w < v : e_w \hb  e_{w+1}$
    \item $\forall w : 1 \leq w \leq v : \newclock.e_w = \newclock.f-v + (w-1)$
    \item $e_{v} \hb  f$

\end{itemize}

Lemmas \ref{lem:lptfnonzeo} and \ref{lem:chain} deal with the case when $\lpt.f$ is greater than $0$. They state that if $\lpt.f$ is greater than $0$ then there is an event $e$ whose (entire) timestamp (i.e., $\newclock.e$) is exactly one less than that of $f$ (i.e., $\newclock.f$). The reverse may, however, not be true. 
%\todo{(Duong: inverse means not p implies not q)}
%
Specifically, it is possible that there exists an event $f$ such that $\lpt.f=0$ and there exists an event $e$ such that $e$ happened before $f$ and $\newclock.e+1=\newclock.f$.
%I.e., it may be possible for $\lpt.f$ to be $0$ and still an event $e$ may exist such that $\newclock.e+1=\newclock.f$ and $e \rightarrow f$. 
Such a situation can arise if the last $\ubits$ bits of $\newclock.j+1$ (or $\newclock.m+1$) happen to be all $0$'s. Essentially, in this case, there is an overflow in the lower $\ubits$ bits thereby affecting the significant bits (i.e., bits that were not considered extraneous). 
With this intuition (and based on Observation \ref{obs:hptincrease}), we define consecutive causal chain. Specifically, a consecutive causal chain is \ccc such that for any $0 < w < \lastccc$,  $e_w \hb  e_{w+1}$, $\newclock.e_w + 1 = \newclock.e_{w+1}$.

\refstepcounter{definitions}\textbf{Definition \thedefinitions}.
\label{def:conscausal}
A sequence of events \ccc is a consecutive causal chain (CCC) iff for each $w, 1 \leq w < \lastccc :$
\begin{itemize}
    \item $ e_w \hb  e_{w+1}$
    \item $\newclock.e_w + 1 = \newclock.e_{w+1}$
\end{itemize}

We note that in a consecutive causal chain \ccc, we cannot insert an event \textit{in between} two events. In other words, we cannot insert event $f$ such that $e_w \hb  f \hb  e_{w+1}$. This is straightforward by Theorem \ref{thm:causality} (cf. Section \ref{sec:logical}) as there is no available timestamp for the event $f$. %\dnote{(This theorem is a little far later. Also I think the reference is Lemma~\ref{lem:chain})}. 
We can only extend a consecutive causal chain by adding an event at the beginning or at the end. 
Hence, we define maximal consecutive causal chain (MCCC) to be the one that cannot be extended further on either end. Note that to extend \ccc, we either need an event $e_0$ such that $\newclock.e_0 = \newclock.e_1 -1$ and $e_0 \hb  e_1$ or an event  $e_{\lastccc+1}$ such that $\newclock.e_{\lastccc + 1} = \newclock.e_{\lastccc}+1$ and $e_\lastccc \hb  e_{\lastccc+1}$.

\begin{comment}
\begin{definition}
A computation of a distributed program is a sequence of states $\br{s_0, s_1, \cdots}$ such that the state $s_i$ is changed to $s_{i+1}$ by some event $e_i$ ($e_i$ could be either a local, send, or receive event), and if $e_i \hb e_j$ then $i < j$. \dnote{(Please help me check this definition which is based on Marzullo~\cite{MN91WDAG}.)}
\end{definition}

\end{comment}

\refstepcounter{definitions}\textbf{Definition \thedefinitions}.
Algorithm \ref{alg:timestamping} has an overflow in assigning timestamps in a computation $\sigma$ 
%A computation of Algorithm \ref{alg:timestamping} has an overflow in a given computation 
iff $\sigma$ contains events $e,f$ such that $e \hb f$ and Algorithm \ref{alg:timestamping} assigns timestamps such that $\newclock.f=\newclock.e+1$, and $\lpt.f=0$.

Note that the above definition is conservative. In other words,   if there is no overflow in the computation it would imply that \hpt is not affected by the increment of \lpt. The reverse is not necessarily true. 
%\todo{S:Rem}
%\snote{Duong: Add the definition of computation}

%\snote{Gabe: We talked about this. But I think this definition is correct. If we have such events that means that there is an overflow because hpt bits are affected by length of the causal chain.}

\refstepcounter{definitions}\textbf{Definition \thedefinitions}.
We say that the deployment of Algorithm \ref{alg:timestamping} in a given system is \textit{\correct} iff for any valid computation (i.e., the one that meets assumptions about $\epsilon$, $\minlocal$, $\minsend, \minreceive$ and $\mintransmit$), there is no overflow (in the assignment of timestamps). 

Next, we focus on identifying the conditions under which Algorithm \ref{alg:timestamping} is a valid timestamping algorithm. 
%Next, we focus on the case where Algorithm \ref{alg:timestamping} is \correct and show that \newclock value is \textit{close} to the physical clock. 
Lemma \ref{lem:chain} will help us determine the limits under which Algorithm \ref{alg:timestamping} can be used. To evaluate the constraints under which the timestamps can be used, consider a CCC \ccc. 
%(Note that the sequence \ref{??} in Figure \ref{??} is an instance of CCC.
%\ccc shown in Figure \ref{??}.
We can make certain observations about the (absolute) time elapsed in creating a given event, say $e_w$

%where $\forall j\  e_j \hb  e_{j+1}$ and there is no event $f$ such that for any $j$, $e_j \hb  f$ and $f \hb  e_{j+1}$. 

%\snote{Add figure here}

\begin{itemize}
    \item If $e_w$ is a local event then the time elapsed is at least $\minlocal$.
    \item If $e_w$ is a send event then the time elapsed from $e_{w-1}$ is at least $\minsend$.
    \item If $e_w$ is a receive event on process $j$ then 
    \begin{itemize}
        \item If $e_{w-1}$ is also on $j$ then the minimum elapsed time is $\minreceive$. (Note that in this case, the message sent at $e_{w-1}$ is unrelated to the message received at $e_w$.) 
        \item If $e_{w-1}$ is not on process $j$ (i.e., event $e_w$ corresponds to receiving the message sent in event $e_{w-1}$) then the minimum elapsed time is $\mintransmit+\minreceive$.
    \end{itemize}
\end{itemize}

Thus, we can observe that for any two events $e_{w-1}$ and $e_{w}$, the elapsed time is at least $min(\minlocal, \minsend, \minreceive)$.
%($\minsend$ for the case where $e_{w+1}$ is a send event, $\minreceive$ for the case where $e_{w+1}$ is a receive event and . 
Thus, the (absolute) time elapsed between $e_1$ and $e_{\lastccc}$ is at least $(\lastccc-1)min(\minlocal,\minsend, \mintransmit)$. Thus, we have 

%Furthermore, by transitivity of happened before for any $w > 1$, $e_1 \hb  e_w$.  By assumption \ref{asm:epsilonexist} the physical clocks differ by at most $\epsilon$. Thus, if $(l-1)min(\minsend, \minreceive, \mintransmit)$ \todo{Gabe: It wasn't immediately obvious to me why this includes $\mintransmit$. It would probably be a good idea to have a footnote explaining why you can't assume that $\minsend$ or $\minreceive$ does not dominate.} exceeds $\epsilon$ then $e_1 \hb  e_l$ must be false. In other words, the length of the chain before the physical clock of $e_l$ exceeds the physical clock of $e_1$ is at most $\frac{\epsilon}{min(\minsend, \minreceive, \mintransmit)}$.  Observe that the value of \lpt increases by exactly 1 in the consecutive causal chain.  Thus, 

\refstepcounter{theorems}\textbf{Theorem \thetheorems}.
\label{thm:nocontamination}
Algorithm \ref{alg:timestamping} is \correct in a given system if
\vspace{-4pt}
$$2^\ubits > 
\left\lceil \frac{\epsilon}{min(\minlocal, \minsend, \minreceive)}\right\rceil $$
\vspace{-4pt}
%\dnote{Formula is good to me}

\textbf{Proof. }
Note that the proof follows from the discussion above except that we need to ensure that for the maximal consecutive causal chain $e_1, e_2, \cdots e_l$, $\lpt.e_1$ must be zero. This follows by letting this be the \textit{first} maximal consecutive causal chain in the computation. 

% \snote{drop this for space}\dnote{(if drop, we need to change the first sentence of implication 2).}
% \refstepcounter{implications}\textbf{Implication \theimplications}.
% Consider a system where nodes are communicating over a gigabit Internet connection. Clocks are synchronized within $10 ms$. 
% %Assuming that we are only timestamping messages (not local events) and $\minsend$ and $\minreceive$ are at least $1 \mu s$
% Assuming  $\minsend, \minreceive,$ and $\minlocal$ are at least  $1\mu s$, 
% (see Appendix Section \ref{sec:measure-minsend-minreceive} for justification), 
% this would give an upper bound of $u=14$. Note that all bits are needed only if a node is exhausting its gigabit connection and sending messages in the worst possible manner for reaching this bound. %\snote{Check that the justification exists.} \dnote{(The Appendix Section~\ref{sec:measure-minsend-minreceive} discusses measuring $\minsend$ and $\minreceive$. Accordingly to Figure~\ref{fig:abstractcomputation}, I can see why $\minlocal$ is greater than $\minsend$ and $\minreceive$. But if we consider local computation as event, than $\minlocal$ could be as smaller than $\minsend$ and $\minreceive$.)}

\refstepcounter{implications}\textbf{Implication \theimplications}.
%Another implication of the above result is that 
The correctness of the timestamping algorithm \textit{can be} independent of the size of the system. We note this is based on the assumption that the value of $\epsilon$ will not increase with the increase in the number of nodes. 
This is reasonable in practice if each node communicates with a clock server directly. 
If nodes synchronize their clocks with each other (with some tree structure to prevent cycles) then the value of $\epsilon$ may depend upon the diameter of the network.  
%This is reasonable, as for most practical systems, the value of $\epsilon$ is likely to be independent of the number of nodes in the system. 
%\todo{Probably need to assume that adding nodes does not increase max physical distance between 2 nodes, since that would affect $\epsilon$}
%\snote{Check above implication rewrite}

\begin{comment}

\refstepcounter{implications}\textbf{Implication \theimplications}.
As an illustration, consider a system that is using only nodes in one AWS region. In this case, clocks can be synchronized to be within \todo{add}. Furthermore, the value of $\minsend/\minreceive/\mintransmit$ are approximately \todo{add}. That means that \todo{add} bits would be sufficient for ensuring that there is no overflow in Algorithm \ref{alg:timestamping}. Or with XYZ number of superfluous bits currently available in Linux, the Algorithm \ref{alg:timestamping} will not overflow if the value of $\minreceive/\minsend/\mintransmit$ is at least XYZ microseconds.
\snote{Duong: can you fix the above implication. See if this needed as well given that $\mintransmit$ is not part of the equation anymore}\dnote{I will add this data}
\end{comment}

Note that Theorem \ref{thm:nocontamination} is only a sufficient condition to use Algorithm \ref{alg:timestamping}. It considers the worst-case scenario where events are created as fast as they can be. And, they can be ordered in the worst possible way. In practice, however, things are likely to be not as pessimistic. For example, processes may not send messages as fast as they can. Message delay may play a role in reducing the length of the longest MCCC.
%consecutive causal chain.

While Theorem \ref{thm:nocontamination} provides only a sufficient condition, it also identifies the key issue that limits conditions under which Algorithm \ref{alg:timestamping} can be used. In particular, it depends upon the length of the longest consecutive causal chain $e_1, e_2, \cdots e_{\lastccc}$ that could occur in absolute time $\epsilon$. 
% \todo{S,D:Rem}
%Specifically, Theorem \ref{thm:nocontamination} assumes that each process sends and receives messages as fast as possible. In other words, there is no internal computation (that takes time) happening on the process. Furthermore, each message is received at the earliest possible time. 
%
To get a more practical estimate about when Algorithm \ref{alg:timestamping} is \correct, we review the consecutive causal chain {(CCC)}. We use the term $\avcomp$ to denote the average (absolute) physical time between two events in the CCC. 
This value can be computed as $\frac{1s}{\textrm{Number of messages sent or received by a process in 1s}}$. 
Furthermore, we use the term $\avtransmit$ to denote average message delay. 
Using the same analysis as in Theorem \ref{thm:nocontamination}, Algorithm \ref{alg:timestamping} should \textit{expect} to work even if 
$$2^\ubits > \left\lceil
\frac{\epsilon}{min(\avcomp, \avtransmit)}\right\rceil \hspace*{15mm} (Equation\ 1)$$
\refstepcounter{implications}\textbf{Implication \theimplications}.
If nodes are transmitting 10K messages per second ($\avcomp = 0.1ms$), the average message delay is $0.25 ms$ (typical latency in AWS servers in the same availability zone) and clock skew is $10 ms$, then only 7 bits would sufficient. 
%\snote{Gabe, Duong: check this} 
%\dnote{(I just used ping to measure. The average round-trip latency within an availability zone is 0.5 ms, between availability zone in same region is 1.4-1.5 ms. So message delay is 0.25 ms instead of 0.5 ms)}
%\todo{Add here by looking at number of messages and message delay}

\subsection{Analysis of \NEWCLOCK for its use as Logical Clock}
\label{sec:logical}

\NEWCLOCK can be trivially used in place of a logical clock \cite{Lamport78CACM} as shown in Theorem \ref{thm:causality}.

\refstepcounter{theorems}\textbf{Theorem \thetheorems}.
\label{thm:causality}
From the algorithm, we observe that if $e$ happened before $f$ then $\newclock.e < \newclock.f$. 

\textbf{Proof. } \ The proof is straightforward and can be proved by induction on the number of events created.

% \vspace*{2mm}

\subsection{Analysis of \NEWCLOCK to be used as Physical Clock }
\label{sec:physical}

We consider two key requirements for \NEWCLOCK to be used in place of physical time: (1) the value of $\newclock.j$ should be \textit{close} to the physical clock of $j$, and (2) the value of $|\newclock.k - \newclock.j|$ should be bounded. We prove this in Lemma \ref{lem:newclockbound} and Theorem \ref{thm:boundeddiff} under the assumption that Algorithm \ref{alg:timestamping} is \correct. For reason of space, the proofs are provided in the Appendix.

\refstepcounter{lemmas}\textbf{Lemma \thelemmas}.
\label{lem:newclockbound}
Consider any computation $\sigma$ where Algorithm \ref{alg:timestamping} does not have an overflow. 
Let $e$ be an event in $\sigma$ on process $j$. Let $\cleanpt.e$ denote the value of $\cleanpt.j$ when $e$ was created. Then,  
\begin{itemize}
    \item $\newclock.e \geq \cleanpt.e$, and
    \item There exists process $k$ such that $\newclock.e \leq \cleanpt.k_e+2^{\ubits}$ where $\cleanpt.k_e$ is the value of $\cleanpt.k$ when $e$ was created.

\end{itemize}

%\snote{Proofs in this section can be moved to appendix}

Note that the above lemma shows an upper bound and lower bound on \newclock.e. In this lemma, the value of $2^{\ubits}$ is very small where $\ubits$ is the number of extraneous bits.  

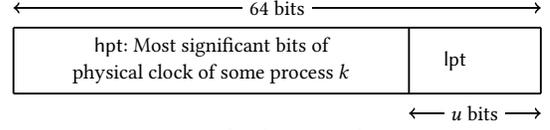
\begin{figure}
\vspace{-14pt}
\centering
% \resizebox{4cm}{1cm}
\resizebox{0.4\textwidth}{!}{
\begin{tikzpicture}
\draw[thick] (0,1) -- (8,1);
\draw[thick] (0,0) -- (8,0);
\draw[thick] (0,0) -- (0,1);
\draw[thick] (6,0) -- (6,1);
\draw[thick] (8,0) -- (8,1);
\draw[thick, <->] (0,1.3) -- (8,1.3);
\draw (4,1.3) node [fill=white] {64 bits};
\draw[thick, <->] (6,-0.3) -- (8,-0.3);
\draw (7,-0.3) node [fill=white] {$u$ bits};

\draw (3,0.7) node {\hpt: Most significant bits of};
\draw (3,0.3) node {physical clock of some process $k$};

\draw (6.7,0.5) node {$\lpt$};
\end{tikzpicture}
}
\vspace{-12pt}
\caption{Format of 64-bit \NEWCLOCK of an event at $j$}
\label{fig:pwc-format}
\vspace{-16pt}
\end{figure}

Based on Lemma \ref{lem:newclockbound}, we can view \NEWCLOCK to be as shown in Figure \ref{fig:pwc-format}. In particular, \newclock is split into \hpt (most significant bits) and \lpt (least significant extraneous bits). The value of \hpt corresponds to the physical clock of some process (but not necessarily the one where the event was created). While it differs from HLC in this regard, we anticipate that given the clock synchronization between physical clock, \newclock can be used in place of HLC.

% \begin{figure}
% \centering
% \begin{tikzpicture}
% \draw[thick] (0,1) -- (8,1);
% \draw[thick] (0,0) -- (8,0);
% \draw[thick] (0,0) -- (0,1);
% \draw[thick] (6,0) -- (6,1);
% \draw[thick] (8,0) -- (8,1);
% \draw[thick, <->] (0,1.3) -- (8,1.3);
% \draw (4,1.3) node [fill=white] {64 bits};
% \draw[thick, <->] (6,-0.3) -- (8,-0.3);
% \draw[thick, <->] (0,-0.3) -- (6,-0.3);

% \draw (7,-0.3) node [fill=white] {\lpt bits};
% \draw (4,-0.3) node [fill=white] {\hpt bits};

% \draw (3,0.7) node {Most significant bits of};
% \draw (3,0.3) node {physical clock of some process $k$};
% %\draw (7,0.5) node {$l-pt$};
% %\draw[thick] (7.5,0) -- (7.5,1);

% %\draw (7.7,0.5) node {$\lpt$};
% \end{tikzpicture}
% \caption{Format of 64-bit \NEWCLOCK of an event at $j$}
% \label{fig:pwc-format}
% \end{figure}

% \gnote{Gabe: This strikes me as a little off. I have a modified version commented below. Which should we use?}

Next, we focus on whether \newclock values themselves are close to each other. Theorem \ref{thm:boundeddiff} shows this result. 

\refstepcounter{theorems}\textbf{Theorem \thetheorems}.
\label{thm:boundeddiff}
If the physical clocks are synchronized to be within $\epsilon$, i.e., at any given time $|pt.j - pt.k | \leq \epsilon$ then at any given time, $|\newclock.j - \newclock.k| \leq \epsilon + 2^{\ubits+1}$ \QEDB
%\dnote{($2^{u+1}$ instead of $2*2^u$?)}

\begin{comment}

\begin{observation}
For any event $e$, $\exists f \clpt.e < clpt.f$
\end{observation}

\textbf{Assumption 1: }
There is at most one \textit{local} event between two \textit{non-local} (send/receive) events). This assumption is reasonable in that if there are multiple events $e_1$ and $e_2$ on process $j$ and $j$ neither sends nor receives a message between these two events then no other process knows the order between them. We can, therefore, assign $e_1$ and $e_2$ the same timestamp. 

\begin{assumption}
\label{asm:epsilonexist}
There exists a bound $\epsilon$ such that the physical clocks of two processes differ by at most $\epsilon$

\end{assumption}

\begin{assumption}
\label{asm:deltasend}
The time required to send a message (in absolute time) is at least $\minsend$.
\end{assumption}

\begin{assumption}
\label{asm:deltatransmit}
The time required to send a message (in absolute time) is at least $\mintransmit$
\end{assumption}

\begin{assumption}
\label{asm:deltareceive}
The time required to send a message (in absolute time) is at least $\minreceive$.
\end{assumption}
\end{comment}

%\subsection{Correctness of Algorithm \ref{alg:timestamping}}

\section{Reliability of \NEWCLOCK} \label{sec:reliable}

The analysis in Section \ref{sec:correct} assumes that all of the assumptions are satisfied. Thus, the natural question is how does \NEWCLOCK behave if some of these assumptions are violated, or if errors occur in the timestamping algorithm. 
%\todo{new text.D:y}
%\snote{We discuss approaches for detecting and correcting errors in \NEWCLOCK next. We also discuss stabilization of \NEWCLOCK in Appendix.}
%\dnote{And remove subsection}

\subsection{Detecting and Correcting Errors in \NEWCLOCK}

As discussed above, the bounds identified in Theorem \ref{thm:nocontamination} are pessimistic. And, in practice, \NEWCLOCK is expected to work correctly even in scenarios where the number of extraneous bits is insufficient to satisfy the conditions identified in Theorem \ref{thm:nocontamination}. 
If \NEWCLOCK is used in scenarios where the condition of Theorem \ref{thm:nocontamination} is violated, there is a potential (however small) that there is an overflow. Thus, a natural question in this context is: \textit{can we detect and correct/eliminate such errors?}

It turns out that the answer to this is already hidden in the proof of Theorem \ref{thm:nocontamination} (cf. Algorithm \ref{alg:detectcorrect}). Here, if there is a possibility of creating an event that leads to a situation where Algorithm \ref{alg:timestamping} is not \correct then we wait to create the corresponding event until the physical time catches up. 
Note that when the physical time catches up, \newclock value is determined by \cleanpt.j instead of \newclock.j or \newclock.m. 
This change also allows us to deal with scenarios where clock skew assumptions are violated. Specifically, we can either delay this message (if the delay was small, e.g., a few milliseconds) or discard the message (if the delay is large several seconds/minutes/more). The former corresponds to errors in clock skew whereas the latter corresponds to malicious clocks/senders or similar errors. The exact cutoff between these choices is application-dependent. 
This choice could also depend upon if the application can tolerate non-FIFO or lost messages. 
%Furthermore, the application would also need to determine if deferring a message should allow subsequent messages to be delivered out of order (if the application is designed to handle messages out of order) or if future messages should be deferred (if the application requires FIFO delivery).

\begin{comment}
\subsection{Changing Clock Skew Over Time}

One can easily imagine a scenario where the value of $\epsilon$ changes over time. In this case, one can easily detect any potential overflow and prevent it before it occurs by adding suitable delays as shown in Algorithm \ref{sec:newclockalgorithm}. 

\end{comment}

%number of bits allocated to \lpt becomes insufficient over time. This may be because $\epsilon$ changes over time as network topology changes, or because protocol changes necessitate more communication events.

%In either case, some strategy to deal with this must be implemented.

\vspace{-10pt}
\begin{algorithm}[ht!]
\small
\SetAlgoLined
\KwResult{Compute $\newclock$ value for send/receive events (Only Send event shown. Change to Receive is similar.)}
\If{process $j$ wants to send $m$}
{
\If{$\newclock.j+1 \mod 2^\ubits = 0 \wedge \newclock.j \geq \cleanpt.j$ }
{
// Either number of extraneous bits is insufficient or clock synchronization assumption is violated or sender's clock is corrupt \;
// Choice 1: Wait until $\cleanpt.j > \newclock.j$ if the wait is small enough \;
// Choice 2: Discard the message if wait is too large (caused by corrupted sender's clock) \;
}
$\newclock.j = max(\newclock.j+1, \cleanpt.j)$ \;
$\newclock.m = \newclock.j$
}

%\If{process $j$ receives a message $m$ with timestamp $l.m$}
%{
%$\newclock.j = max(\newclock.j+1, \newclock.m+1, cleannpt.j)$ \;
%}
 \caption{Detecting and Correcting Errors in \NEWCLOCK}
 \label{alg:detectcorrect}
\end{algorithm}
\vspace{-10pt}

\subsection{Dealing with Transient Errors in \NEWCLOCK}

%move to section 6

%Stabilization refers to the ability of the system to recover from an arbitrary state to a legitimate state \cite{EDW426}. Stabilization guarantees that recovery from arbitrary transient faults, i.e., faults that are temporary even if their cause is unknown/unexpected. 

%It is unclear if \NEWCLOCK would be stabilizing using the Algorithm \ref{alg:detectcorrect} even if we assume that NTP clocks themselves are stabilizing. 
Given the long term usage of NTP in practice, it is anticipated it recovers from transient faults encountered in practice. Now, we consider whether \NEWCLOCK can recover from transient errors as well. 

%While the stabilization of NTP clocks is not formally known, they are \textit{practically stabilizing}. By \textit{practically stabilizing}, we mean that it has been widely used in various environments and ensures that clocks of processes remain close to each other without human intervention. Thus, it is recovering from the vast majority of states that the system could be perturbed to. 

%A natural question is whether \NEWCLOCK will stabilize, assuming that NTP clocks do.

We note that Algorithms \ref{alg:timestamping} and  \ref{alg:detectcorrect} will not recover from transient faults on their own even if NTP clocks do. Specifically, there are two problems: (1) Perturbation of \newclock could perturb it to a really large value. And, Algorithms \ref{alg:timestamping} and \ref{alg:detectcorrect} do not have a mechanism to reduce \newclock value, and (2) It may be possible that \lpt bits are very high when NTP clocks recover from transient faults thereby causing \NEWCLOCK to have repeated overflow after NTP clocks recover from transient faults. 
%It is unclear if this overflow can keep on happening forever after NTP clocks have stabilized. 

The first problem can be solved with Theorem \ref{thm:boundeddiff}. Specifically, if $\newclock.j$ is outside the range $[\cleanpt.j, \cleanpt.j+\epsilon+2^\ubits]$ then $\newclock.j$ should be reset to $\cleanpt.j$.
The simplest way to solve the second problem is to use the notion of reset \cite{AG94TC}; specifically, when an overflow is detected, the detecting process can request all others to suspend any communication in time $[t, t+2\epsilon]$, where $t$ is some time chosen in the future. This will guarantee that $\lpt$ values are $0$ when the computation restarts.

While simple, the above approach is intrusive; it requires the application to stop sending messages for a $2\epsilon$ window. This overhead can be reduced by requiring a process to initiate reset only after a threshold number of instances where there is a likely overflow.

\section{Experimental Behavior of \NEWCLOCK} \label{sec:experiments}

To evaluate the number of extraneous bits required for \NEWCLOCK, we performed simulations via a discrete event simulator and experiments on an actual network. We present these results, next.
\subsection{Simulation Results}
%Duong: should we need a symbol for message send rate to make notions consistent?
To analyze the behavior of \NEWCLOCK, we implemented a discrete event simulator that takes $\epsilon, \minsend, \mintransmit, \minreceive$, and message rate as parameters. 
In this simulator, one clock tick corresponds to $1 \mu s$. Processes advance their clocks subject to the constraint that maximum clock skew remains within $\epsilon$. At each clock tick, the process receives any messages meant to be received at that clock tick. It also may send a message (based on message rate) whose delivery time is set up by the message latency. Source code and raw data are available at \cite{Appleton2020PWCData}. %\url{https://gist.github.com/AnonymousPaperPWC/39001b43cda3b5aa3a99783b0b418c74}.

We let $\minsend$ to be within $1\!-\!12\ \mu s$, $\minreceive$ to be within $1\!-13\ \mu s$, message latency to be within $1-20\ ms$, $\epsilon$ to be $6.25 ms\!-\!400 ms$, 
% message rate ($1K\!-\!64K$ messages per second) messages per node per second), 
message rate to be within $1K\!-\!64K$ messages per node per second, 
and the number of nodes between $8\!-\!64$. We consider three types of networks: (1) a random network where each node communicates with every other node and message destination is selected from uniform distribution, (2) a time leader network where the clock of one node is consistently ahead of other nodes in the network, and (3) a hub and spoke network (client-server network) where we have one hub node (server) and multiple spoke nodes (clients).
%is a server and all clients communicate with it. 
Each simulation is run for approximately 1000 simulated seconds, generating several million events at a minimum. 
Since the data for 32 and 64 nodes does not add a significant value for discussion, we omit it. However, the raw data is available at \cite{Appleton2020PWCData}.

\begin{figure}[t]
    \centering
    \vspace{-16pt}
    \includegraphics[width=\linewidth]{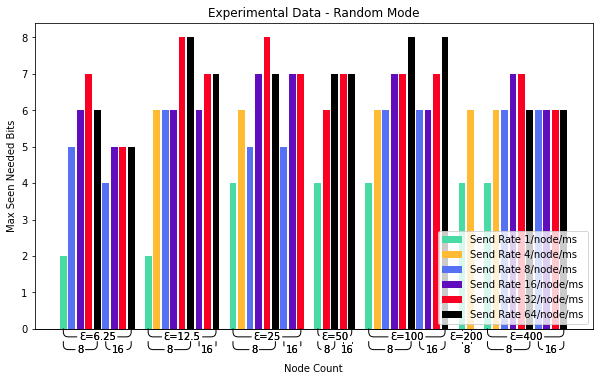}
    \vspace{-20pt}
    \caption{Value of \ubits for wait-free \correctness (Random Network)}
    \label{fig:exp_results_random}
    \vspace{-14pt}
    % \vspace{-10pt}
\end{figure}

\begin{figure}[t]
    \centering
    \vspace{-16pt}
    \includegraphics[width=\linewidth]{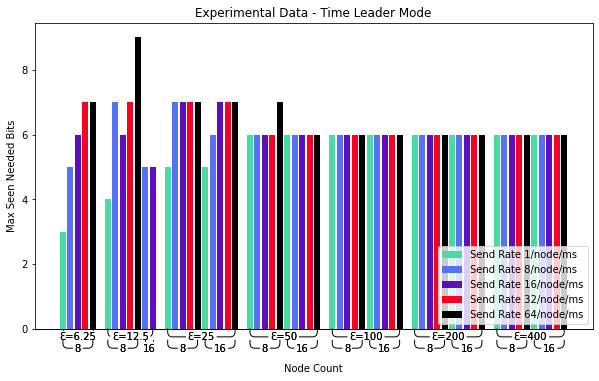}
    \vspace{-20pt}
    \caption{Value of \ubits for wait-free \correctness (Time Leader Network)}
    \label{fig:exp_results_leader}
    \vspace{-14pt}
\end{figure}

\begin{figure}[t]
    \centering
    % \vspace{-16pt}
    \includegraphics[width=\linewidth]{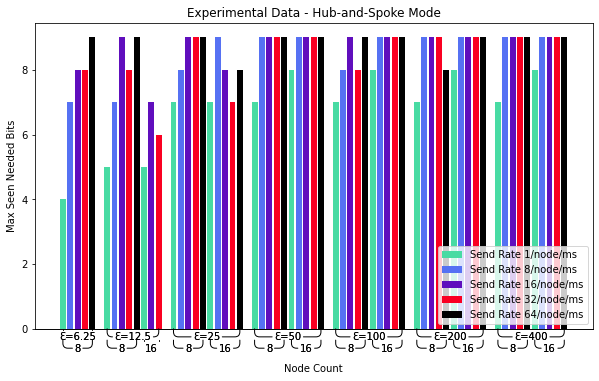}
    \vspace{-20pt}
    \caption{Value of \ubits for wait-free \correctness (Hub-and-Spoke Network)}
    \label{fig:exp_results_hub}
    \vspace{-14pt}
\end{figure}

All networks were simulated with uniform traffic rather than burst traffic. While this is less realistic, it represents a worst-case scenario for the algorithm, as unlike in a burst mode the network will not have time to recover from high-traffic scenarios. Additionally, if a network has a regular traffic of 1K messages per second per node and a burst of 8K messages per second per node then the number of extraneous bits required is at most equal to the case where there is a regular traffic of 8K messages per node per second. The results are shown in Figures \ref{fig:exp_results_random}, \ref{fig:exp_results_leader}, and \ref{fig:exp_results_hub}. From these simulations, we find that the number of bits needed is 9 or less (which corresponds to applications requiring resolution of 120 $ns$). The median is less than 6 (which corresponds to applications requiring resolution of $15 ns$).%\snote{check} \dnote{(I think $15 ns$ for $u=6$)}

We note that the analysis in Section \ref{sec:correct} is conservative in nature. Hence, we use the simulation data to compare the bounds from Section \ref{sec:correct} and simulated observations. Specifically, we find that the value of $\ubits$ is given by the following formula.
\vspace{-3pt}
\begin{samepage}
$$\ubits \approx \left\lceil\frac{\left(\log_2\left(\frac{1000 \cdot S^2}{\min(\minreceive, \minsend)}\right)+\frac{\log_2\left(\epsilon\right)}{\log_{2}\left(S+1\right)}\right)}{K}\right\rceil$$
Where $K$ is a weighting constant we find to be $2.9 \pm 0.1$, $S$ is the send rate in messages/node/millisecond,  $\epsilon$ is the max clock skew in milliseconds, and $\minreceive, \minsend$ are in microseconds. Based on this, we find that the actual value of $\ubits$ is roughly one-third of that predicted by Equation 1 in Section \ref{sec:correct}.
\end{samepage}
%In all simulations run, the largest number of bits needed to prevent overflow was 9. The median was 6, with a mean of $\sim \! 5.88$.

% \todo{S: Rem}

\begin{comment}

\subsection{Random Network}

A random network represents the baseline case for this algorithm. In this mode, the number of bits needed can be predicted by the following formula: $$\ubits \approx \left\lceil\frac{\left(\log_2\left(\frac{1000 \cdot S^2}{\min(\minreceive, \minsend)}\right)+\frac{\log_2\left(\epsilon\right)}{\log_{2}\left(S+1\right)}\right)}{K}\right\rceil$$

Where:
\begin{itemize}
    \item $K$ is a weighting constant we find to be $2.9 \pm 0.1$
    \item $S$ is the send rate in messages/node/millisecond
    \item $\epsilon$ is the max clock skew in milliseconds
    \item $\minreceive, \minsend$ are in microseconds
\end{itemize}

\end{comment}

Additionally, we make the following observations.
%\dnote{(I have difficulty in connecting the observation with the data/figures.)}

\subsubsection{Effect of $\epsilon$, clock skew. }\hfill\newline
Clock skew is the dominant effect in a random network, and it is logarithmic in nature. In particular, for every $K$ ($\approx 3$) doublings of $\epsilon$, one should expect to need 1 extra bit.

\subsubsection{Effect of message rate. }\hfill\newline
Increasing from a low message rate (1 msg/node/ms) to a moderate one (4 msg/node/ms) has a large impact, but it drops off quickly after that. In particular, higher send rates seem to dilute the importance of clock skew while increasing the baseline number of bits needed.
% \todo{S:Rem}
%In graphical terms: if \ubits is the y axis and $\log_2(\epsilon)$ the x axis, increasing $S$ will move the x-intercept to the left while decreasing slope.

\subsubsection{Effect of the number of nodes. }\hfill\newline
If increasing the number of nodes in this type of system has an impact, we were unable to observe it in random network mode. This is predicted by Equation 1 which is independent of the number of nodes.
This also makes sense, because a node should expect to see 1 in (N-1) messages from other nodes, and there are (N-1) other nodes sending messages, so these should cancel out as observed. %\dnote{(I do not understand the last sentence. I am not sure if we are saying the chance of receiving a message is independent of the number of node when the send rate is unchanged?)}

\subsubsection{Effect of network topology. }\hfill\newline 
We note that the results for the leader network mode and hub and spoke network mode are similar with subtle differences. Specifically, in the leader network mode, the actual clock skew is higher as one node consistently is ahead of others. This causes a higher baseline as well as larger variation. The same effect is observed in hub and spoke network where the hub node sees all the messages from the spoke nodes. For the hub and spoke network, we also observe that the value of $\ubits$ is more sensitive to $\epsilon$ especially at low values. For both these networks, we observe higher variability in the number of needed bits resulting in certain anomalies such as those where increasing the value of $\epsilon$ causes the value of $\ubits$ to reduce.

\subsubsection{Causes of anomalies in simulation results. }\hfill\newline 
We note that we do see some exceptions to the above observations. We note that these are due to the long tail of a probabilistic distribution where a very small number of events need a larger number of bits. As an illustration, we consider the result for $\epsilon=6.25$ms, $N=8$, $S=64$. where $\sim\!736$M events were created.

Of these, $\sim\!731$M events needed 0 bits, $\sim\!2$M events needed 1 bit. Only 142 events needed 9 bits (the maximum observed for this simulation). In other words, less than 1 in 1 million events needed all 9 bits. %\dnote{(maximum number of bits could be misleading)}. 
Since the probability of occurrence of such events is very low, we do see some instances where even with a small value of $\epsilon$, the number of bits required is high.

We see that this effect is higher in hub-and-spoke network; this is partly due to the fact that if the hub node generates an event that needs a higher number of bits then it has a substantial potential to affect multiple spoke nodes. In addition, the hub node will see events generated by every other node, and thus has a higher than normal potential to see high-\lpt events.
{The raw data for the number of events and bits required is available at \cite{Appleton2020PWCData}}.

%\snote{put a table rather than figure} \gnote{commented table now has correct data}

\subsubsection{Delays due to insufficient extraneous bits. }\hfill\newline
The above analysis also shows that if the number of available bits is low and we choose to delay certain messages (as done in Algorithm \ref{alg:detectcorrect}) then the number of affected messages is very small. For example,  in the analysis from the previous paragraph, even though 9 bits are required for wait-free operation, if we had used only 4 (respectively 6) bits then 0.033\% (respectively 0.01\%) of messages would be delayed. 
%\dnote{(I'm good with this section. I'm just a little concerned that the discussion text is a not very connected with the Figures. What I mean is we could say, in Figure A, column B, etc. we observe that.)}
\begin{comment}

\begin{tabular}{|c|c|c|c|c|c|c|c|c|c|c|}
\hline
0 & 1 & 2 & 3 & 4 & 5 & 6 & 7 & 8 & 9\\
\hline
730,857,940	& 2,087,980 &	1,672,314 &	936,200 &	247,388 &	68,788 &	78,028 &	24,432 &	3,278 &	142\\
% 15M & 200K & 10K & 2K & 1K & 143\\
\hline
\end{tabular}
\end{comment}

\begin{comment}

\begin{figure}[h]
    \includegraphics[width=0.9\linewidth]{Images/BitsNeededBar.E6250.N8.S64.H.png}

    \caption{Log-scale number of events needing a given number of bits for the simulation of $\epsilon=6.25$ms$,$ $N=8,S=64$ in Hub-and-Spoke mode}
    \label{fig:BitsNeededBar.E6250.N8.S64.H}
\end{figure}
\end{comment}

\subsection{Experimental Analysis}
\label{sec:experimentalanalysis}

In addition to the above, we performed a version of this experiment on physical hardware. This allows less control over the environment, but produces about 100 times the number of events in a similar time span.

The experiment was performed using seven hosts on a local network, whose send rates varied from $\sim\!19,000$ messages sent per second to $\sim\!126,000$ at the other extreme. The average send rate across the system was $\sim\!43,000$ messages per second. Traffic was generated randomly, as in the simulation's random mode, except that it was not rate limited. Messages were sent as fast as the hardware allowed while maintaining consistent locking on the \pwc object.

\begin{figure}[t]
    \centering
    \vspace{-16pt}
    \includegraphics[width=\linewidth]{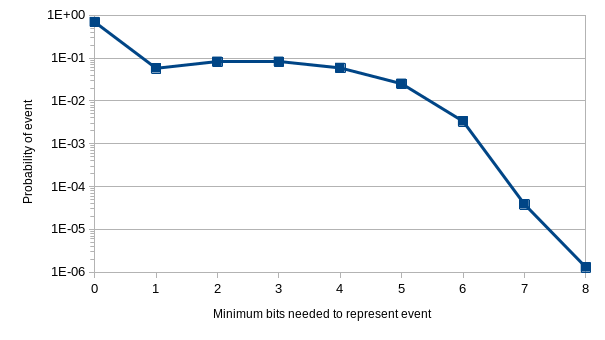}
    \vspace{-24pt}
    \caption{Probability of needing a given number of bits to represent a timestamp without delay}
    \label{fig:exp_raw_prob}
    \vspace{-14pt}
\end{figure}

\begin{figure}[t]
    \centering
    % \vspace{-16pt}
    \includegraphics[width=\linewidth]{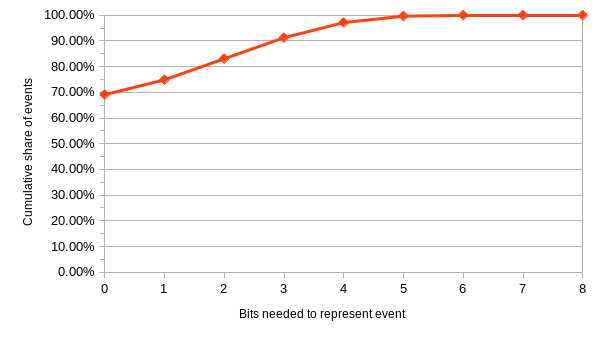}
    \vspace{-24pt}
    \caption{Given number of bits, what portion of events can be represented by a timestamp without delay}
    \label{fig:exp_cum_percentage}
    \vspace{-14pt}
\end{figure}

Overall, the experiment had results very similar to the simulations. About 1 in 25,000 events required more than 6 extraneous bits, and about 1 in 770,000 required more than 7. All events of the $\sim\!103$ billion could be represented with 8 extraneous bits. From this, we find that the number of bits needed is 8 or less (which corresponds to applications requiring resolution of 60 $ns$).

\begin{figure}[t]
    \centering
    \vspace{-16pt}
    \includegraphics[width=\linewidth]{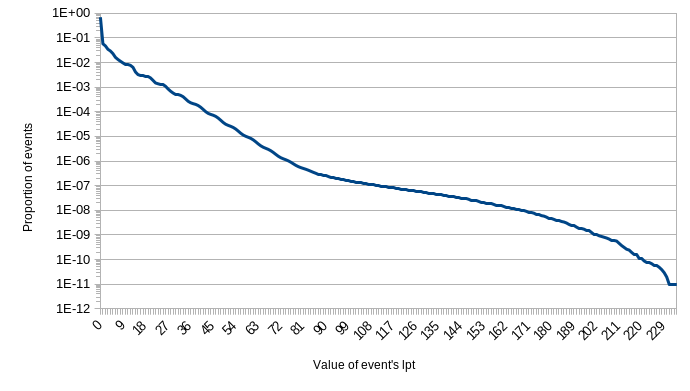}
    \vspace{-24pt}
    \caption{Proportion of events with a given $\lpt$ value}
    \label{fig:exp_indv_prob}
    \vspace{-14pt}
\end{figure}

One feature of Figure \ref{fig:exp_raw_prob} which may seem unusual at first glance is that the number of events with an $\lpt$ bit width of 1 is smaller than of sizes 0 and 2. This makes sense, however, when one considers the number of values covered by that range. Bit width 0 can only represent the value 0, so all timestamps with a freshly generated $\hpt$ will generate this. As seen in Figure \ref{fig:exp_indv_prob}, events with $\lpt=0$ are very common. Moving to a bit width of 1 only allows you to represent one additional value. So while values in bit widths 2, 3, and 4 are less common, this is overcome by the fact that they add a larger range of possible values.

As above, all data and source code may be found at \cite{Appleton2020PWCData}.

\section{Discussion}
\label{sec:discussion}
In this section, we discuss some of the questions raised by \NEWCLOCK. A few additional questions are discussed in Appendix. 

\subsection{Comparison of HLC and \NEWCLOCK}

There are some subtle differences between the clocks presented in this paper and those in \cite{KDMA2014OPODIS.HLC}. 
%These subtle differences will illustrate how HLC differs from \NEWCLOCK. 
HLC maintains two parts: $l$ and $c$, where $l$ value is close to the physical clock and $c$ is a counter. To make it backward compatible with the physical clock, the extra information associated with HLC is saved in the lower bits of the physical clock. 
The most significant bits of the HLC timestamp for event $e$ on process $j$ correspond to the physical time of $j$ when event $e$ occurred. By contrast, from Lemma \ref{lem:newclockbound}, the most significant bits of \NEWCLOCK timestamp for event $e$ on process $j$ corresponds to the physical time of \textit{some} process $k$ in the system. 
%For example, Given an event with physical clock $100$, $l=102$ and $c=3$, this information is stored in HLC as shown in Figure \ref{fig:??}. Specifically, the lower bits of physical clock saves the value of $l-pt$ and $c$. By contrast, with \newclock, it would be saved as shown in Figure \ref{??}. 

\subsection{Disadvantages of \NEWCLOCK}
These subtle differences have several implications: For one, the most significant bits of HLC of event $e$ on process $j$ match with the physical clock of $j$.
%those corresponding to the physical clock. 
This means that if HLC is used to compare time on the same process then the difference in HLC value is expected to be \textit{close} to that in the physical clock. On the other hand, in \NEWCLOCK, the most significant bits correspond to the physical clock of \textit{some} process but not necessarily process $j$.
%represent the logical clock. 
This means that if we compare two \newclock values on the same process then their difference may be affected by $\epsilon$.

% \todo{S: Rem}
%\snote{Remove? 
%By the same token, the HLC timestamp of event $e$ on process $j$ can be used to identify the physical time of $j$ when $e$ occurred. By contrast, \NEWCLOCK provides physical clock of some process (not necessarily $j$)
%}

%Also, the way HLC stores information, it is possible to extract both the physical clock (most significant bits) as well as the complete HLC timestamp ($l$ and $c$ value) from the information stored in HLC. By contrast, with \NEWLOCK, one cannot obtain the physical clock value of the event. 

\subsection{Advantages of \NEWCLOCK over HLC}
The above benefits of HLC come at a cost that makes \NEWCLOCK attractive in many scenarios. For example, in HLC, the difference between the $l$ value and the $pt$ value can be as large as $\epsilon$, the clock skew. Thus, if the clock skew is $10ms$ then the bits allocated to $l-pt$ must be large enough to represent a skew of $1ms$. If clock granularity is $10 \mu s$ then it implies that $l-pt$ may have 1000 possible values thereby requiring 10 bits. If the clock granularity is $0.1 \mu s$ and clock skew is $10 ms$ then $l-pt$ would require 17 bits. This is a severe limitation for HLC. (Plus, bits required for $c$ are extra.)
In \NEWCLOCK, however, the value of $l-pt$ is not stored in the lower bits of the physical clock. For example, if the clock skew is $10ms$ and each process sends 1000 messages per second then (based on the analysis in Section \ref{sec:correct}), having just 4 extraneous bits is sufficient. Thus, \NEWCLOCK is useful in many scenarios where HLC is not. 

HLC timestamp puts limits under which HLC can be used. In particular, if 16 bits are no longer available, it implies that the application must not generate two events within $15\ \mu s$. As $\minsend$ and $\minreceive$ could be as small as $1\ \mu s$, this resolution is not sufficient if an application needs to send several messages as fast as possible in a short period of time (micro-burst). By contrast, \NEWCLOCK can be used even in scenarios where an application needs clock resolution of $10-100ns$. 

Additionally, \NEWCLOCK is more resilient to clock skew errors. If clock skew errors increase, it may not be possible to save $l-pt$ (whose value can be as large as the clock skew) in the assigned space. By contrast, \NEWCLOCK is not affected by transient clock skew errors. As discussed in Section \ref{sec:experiments}, even a clock synchronization error of $400 ms$ does not significantly affect the number of extraneous bits required. %\snote{plese see above two para}

Another, possibly more important benefit of \NEWCLOCK is that the comparison of clocks is just an integer comparison. It does not require decoding the timestamp necessary for HLC.
If we just use integer comparison then the comparison of HLC timestamp can provide incorrect information.

\vspace*{-3mm}
\subsection{HLC Limitation for Comparison of Timestamps}
\label{sec:HLClimitations}

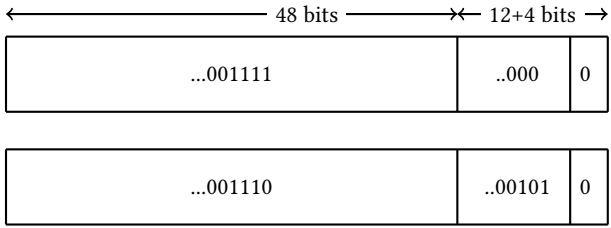
\begin{figure}[h]
\centering
\vspace{-8pt}
\begin{tikzpicture}
\draw[thick, <->] (0,1.3) -- (6,1.3);
\draw (4,1.3) node [fill=white] {48 bits};
\draw[thick, <->] (6,1.3) -- (8,1.3);
\draw (7,1.3) node [fill=white] {12+4 bits};
% event e
\draw[thick] (0,1) -- (8,1);
\draw[thick] (0,0) -- (8,0);
\draw[thick] (0,0) -- (0,1);
\draw[thick] (6,0) -- (6,1);
\draw[thick] (8,0) -- (8,1);
\draw (3,0.5) node {...001111};
\draw (6.8,0.5) node {..000};
\draw[thick] (7.5,0) -- (7.5,1);
\draw (7.7,0.5) node {0};
% event f
\draw[thick] (0,-0.5) -- (8,-0.5);
\draw[thick] (0,-1.5) -- (8,-1.5);
\draw[thick] (0,-1.5) -- (0,-0.5);
\draw[thick] (6,-1.5) -- (6,-0.5);
\draw[thick] (8,-1.5) -- (8,-0.5);
\draw (3,-1) node {...001110};
\draw (6.8,-1) node {..00101};
\draw[thick] (7.5,-1.5) -- (7.5,-0.5);
\draw (7.7,-1) node {0};

\end{tikzpicture}
\vspace{-6pt}
\caption{Example of HLC timestamps of events $e$ (top) and $f$ (bottom). By HLC format, the timestamp of $e$ is smaller than $f$. However, if the timestamps are treated as 64-bit integers, the value for $e$ is larger than the value for $f$. (Leading zeros are omitted)}
\label{fig:hlc-encode-decode}
\vspace{-10pt}
\end{figure}

In this section, we discuss the claim that HLC timestamps cannot be compared with standard $<$ operation to deduce causal information. To illustrate this, consider two timestamps $e$ and $f$ with HLC timestamps shown in Figure \ref{fig:hlc-encode-decode}.
Observe that with these timestamps $l.e=15, c.e=0$ and $l.f=14+5=19, c.f=0$. However, if we compare these timestamps as integer timestamps then we would conclude that $HLC.e$ is larger than $HLC.f$. In other words, without suitable encoding/decoding, a comparison of HLC timestamps would lead to incorrect conclusions. By contrast, \NEWCLOCK does not have this issue. 
In other words, 
backward compatibility of \NEWCLOCK is superior to that of HLC.

%\vspace{-7mm}

\vspace*{-3mm}

%This implies that even if Furthermore, while both HLC and \NEWCLOCK are backward compatible with NTP clocks (since they use lower end bits of the clock), \NEWCLOCK 
\subsection{Dealing with an Unknown Value of $\epsilon$ }
%For simplicity, we assumed that the value of $\epsilon$ is known to processes. However, if you observe the code in Algorithm \ref{alg:timestamping}, the value of $\epsilon$ is never used. 
%Algorithm \ref{alg:timestamping} only used the value of $\ubits$; this value is determined by the time resolution needed for the application and not the value of clock skew. 
%
Algorithm \ref{alg:timestamping} did not use  $\epsilon$.
Hence, the algorithm can be used even if $\epsilon$ is not known or is highly variable. %
Value of $\epsilon$ is needed only to determine if Algorithm \ref{alg:timestamping} is \correct. Furthermore, the changes in Algorithm \ref{alg:detectcorrect} can be used to ensure that the algorithm remains \correct by adding occasional small delays on messages. Thus, \NEWCLOCK can be easily used in scenarios where the exact value of $\epsilon$ is not known.

\subsection{Dealing with Dynamic Value of $\ubits$}
\label{sec:dynamicu}
The value of $\ubits$ can also be changed dynamically. The easiest way to do this is to communicate to all processes that the value of $\ubits$ should be changed to $\ubits'$ at \NEWCLOCK time $t'$, where $t'$ is sufficiently larger. Furthermore, it is also possible that $\ubits$ can be changed independently. For example, consider the case where process $j$ assumes $\ubits=4$ and process $k$ assumes $\ubits=5$. In this case, if Algorithm \ref{alg:timestamping} remains \correct under the assumption that $\ubits=4$ then it would be correct under the scenario where process $j$ treats $\ubits=4$ and process $k$ treats $\ubits=5$. {Due to reasons of space, we omit the proof of this claim}

\subsection{Dealing with Non-Monotonic Clocks}
\label{sec:nonmonotonic}

While a monotonic clock is ideal for running \NEWCLOCK, it can handle a transient clock drift due to (for example) a negative leap second.
First, we observe that this type of underlying clock drift does not affect causality, i.e., even in the presence of a negative leap second, Theorem \ref{thm:causality} remains valid. It can only affect the condition related to the overflow of the clock. Specifically, if a clock can go backward, the number of events that may be created for a given \hpt value increases. In turn, the number of bits required to prevent overflow may increase. 
Furthermore, the approach used in Algorithm \ref{alg:detectcorrect} will handle potential overflows by delaying events when needed. 
%So long as the timestamping algorithm does not overflow before the underlying clock catches up to \NEWCLOCK, the algorithm should recover without error or message delay.

\subsection{Use of \NEWCLOCK in Evaluating Elapsed Time}
%\label{sec:elapsed}

One use of physical clocks is in evaluating time elapsed between two events. Specifically, if $pt.e$ and $pt.f$ denote the physical clock of event $e$ and $f$, we can  use $pt.f-pt.e$ to determine the time elapsed between $e$ and $f$.

%If events $e$ and $f$ are on the same process then in the absence of NTP correction, this would be \textit{close} to the actual physical time.  

Both HLC and \NEWCLOCK can be used as a substitute. Errors due to the use of HLC or \NEWCLOCK are comparable in nature. In HLC, the $l$ value can differ from $pt$ by $\epsilon$ (with $l$ value always higher or equal). Hence, while comparing $l.f-l.e$, there is a potential error of $2\epsilon$ when compared with $pt.f-pt.e$. A similar error can occur with \NEWCLOCK.

%When computed with absolute physical time, $pt.f-pt.e$ can have an error of $2\epsilon$, as $pt.f$ could be ahead of $\epsilon$ with respect to absolute physical time, and $pt.e$ could be behind by $\epsilon$ with respect to the absolute physical time. Error in HLC or \NEWCLOCK is identical. In this sense, \NEWCLOCK (as well as HLC) provide the same level of accuracy for computing elapsed time as physical clock itself. In other words, using the extraneous bits of physical clock to obtain \NEWCLOCK has preserved the ability to compute elapsed time with \NEWCLOCK. 

\subsection{Use of \NEWCLOCK as Logical Clock}

Based on Theorem \ref{thm:causality}, we see that \newclock can be used in any place where logical clock can be used. For example, it can be used in mutual exclusion algorithms such as Lamport's algorithm \cite{Ghosh14}. 
% \todo{add more}

\subsection{Use of \NEWCLOCK as Physical Clock}

\NEWCLOCK can also be used as a replacement for physical clocks. Specifically, algorithms such as Spanner rely on physical clocks (synchronized within some bound $\epsilon$). Based on Theorem \ref{thm:boundeddiff}, \NEWCLOCK can be used in its place. Furthermore, with \NEWCLOCK, some of the deficiencies (e.g., commit-wait delay) are removed without adding any overhead. Likewise, \NEWCLOCK can also be used in algorithms such as CausalSpartan \cite{CausalSpartan.RDK17SRDS} to obtain causal consistency without introducing delays caused by clock skew. As discussed above, HLC can also be used to remove some of the disadvantages discussed above. However, a key advantage of \NEWCLOCK over HLC is that \NEWCLOCK can be used even when clock skew is high. In this case, HLC is limited as bits are needed to save the $l-pt$ value.
By contrast, \NEWCLOCK can be used even with large clock skews as it does not explicitly store values such as $l-pt$. 
%\snote{how about now?}
%\dnote{(I think \NEWCLOCK is still affected by clock skew $\epsilon$ as shown in Theorem~\ref{thm:nocontamination}, is that correct?)}

\section{Related Work}
\label{sec:related-work}

% Examining or debugging a distributed computation relies on execution trace of events which are labeled by the moment they occurred.
% The desirable properties for the timestamp are accuracy (the timestamp help to identify the order among events correctly) and cost-effective.
% There are three approaches for timestamp events in a distributed computation.
% Physical clock based approaches. Using physical clock attached to each computer machine is a natural, efficient and common solution to record event timestamps. However it suffers from low accuracy. Due to clock skew associated with clock synchronization protocol~\cite{ntp}, it is possible that $pt.e > pt.f$ but $e$ actually occurred after $f$ or vice versa, i.e. $e$ actually occurred after $f$ but $pt.e > pt.f$. True time~\cite{CDEFFFGGHH13TOCS} removes this drawback but it requires dedicated hardware and if there is any uncertainty in the ordering of events, it delay the computation to clear the uncertainty.
% Logical clock based approaches: Lamport clock, logical clock, vector clock, Kronos system, dependency graph.
% Combination of both.

Recording timestamps of events is critical for reconstructing and debugging the execution of a distributed program. 
The natural solution to record timestamps of events is the physical time which is the reading of the local clock of the machine on which the process runs. However, physical clocks are not perfectly synchronized, which makes
it possible that $pt.e > pt.f$ but $e$ actually occurred after $f$ or vice versa, i.e. $e$ actually occurred after $f$ but $pt.e > pt.f$.

TrueTime~\cite{CDEFFFGGHH13TOCS} is another physical-clock-based approach but it requires dedicated hardware as well as delayed operation if needed to clear the uncertainty. 
Lamport's logical clock~\cite{Lamport78CACM} guarantees that if event $f$ causally depends on event $e$ then the logical clock of $f$ is strictly greater than $e$. However, the converse of this property is not true, i.e. one could not infer the causality between two events based on their logical timestamps. 
In \cite{fetzer}, authors use partially synchronized clocks to implement several protocols in distributed systems. 
The concept of encoding low-end bits to provide reliable ordered delivery is considered in \cite{sscmp}. 
The two-way causality property is provided by vector clocks~\cite{Mattern89PDA,Fidge87}, dependence blocks~\cite{SM94DC} which keep track of and combine causality information on all processes. However, the size of vector clocks is $O(n)$ where $n$ is the number of processes, which is prohibitively expensive for large distributed systems. To handle the dynamic nature of distributed systems and reduce the space complexity of vector clocks, interval tree clocks~\cite{ABF2008OPODIS.intervaltreeclock} and Bloom clocks~\cite{ramabaja2019arxiv.bloomclock} are proposed.

Both logical and vector clocks are timestamps that completely independent from the physical clock. Some works combine the information of logical and physical clock to improve the causality decision as well as reduce the size of the timestamp~\cite{KDMA2014OPODIS.HLC,DK13LADIS,VK2018SIROCCO}. Among those, Hybrid Logical Clock~\cite{KDMA2014OPODIS.HLC} is the work closest to the \pwc described in this paper. In HLC, the bit array storing the timestamp is separated into two parts: one for physical clock information and the other for logical clock information (causality). Thus, manipulation of HLC requires encoding and decoding the bit array. By contrast, in \pwc, the whole bit array is treated as a single integer and there is no need for encoding/decoding the causality information.

\vspace*{-3mm}
\section{Conclusion}
\label{sec:concl}

In this paper, we presented \NEWCLOCK that was a physical clock that also provided information to deduce one-way causality.  We achieved this by observing that a certain number of extraneous/redundant bits are available in a physical clock. We used these bits to provide causal information. 

%\snote{Check new text}
\NEWCLOCK provides several benefits over HLC \cite{KDMA2014OPODIS.HLC}. 
For one, \NEWCLOCK is applicable in many systems where clock skew is larger or more variable. 
%\dnote{susceptible to transient spikes.}
HLC is limited by the fact that it needs to store the value of $l-pt$ and this value depends upon $\epsilon$, the clock skew. By contrast, \NEWCLOCK is unaffected by transient errors that cause significantly higher clock skew. 
Second, \NEWCLOCK works correctly even if the number of extraneous bits is small; we find that even 9 bits (time resolution of 120 $\ ns$) are sufficient for \NEWCLOCK. By contrast, HLC generally requires a larger number of bits (16 proposed in \cite{KDMA2014OPODIS.HLC} which corresponds to clock resolution of $15 \mu s$). Hence, HLC is not able to handle microbursts of messages that \NEWCLOCK can handle. Thus, \NEWCLOCK is more applicable than HLC. %
Third, the backward compatibility of \NEWCLOCK with physical clocks is better than that of HLC. Specifically, to get the HLC comparison right to deduce (one-way) causality, we need to decode the timestamps. With \NEWCLOCK, just integer comparison suffices. 

A key disadvantage of  \NEWCLOCK over HLC \cite{KDMA2014OPODIS.HLC} is that the most significant bits of $\newclock.{e_j}$ correspond to the physical clock of some process $k$. By contrast, the most significant bits of  $hlc.{e_j}$ correspond to the physical clock of $j$.
% \todo{S: Rem}
%This means that the  difference  in the \NEWCLOCK values of consecutive events on the same process \textit{may} be affected by clock skew in the system. By contrast, with HLC, the difference between two consecutive events on the same process does not depend upon clock skew. 

HLC and \NEWCLOCK provide several common features. Both are strictly increasing. Both can be used in place of logical or physical clock. Both can be used to eliminate delays caused by clock skew in applications such as Spanner \cite{CDEFFFGGHH13TOCS}, CausalSpartan \cite{CausalSpartan.RDK17SRDS}.

There are several opportunities provided by \NEWCLOCK that would permit additional future extensions of \NEWCLOCK. For example, 
\NEWCLOCK also provides a possibility of obtaining (two-way) causality with physical clocks. Specifically, if the available extraneous bits exceed the number of bits needed to prevent overflow, additional bits could be used to provide additional information that could be used to deduce when two events are concurrent. 

\vspace{-4pt}
\bibliographystyle{IEEEtran}
\bibliography{main}

\newpage

\appendix
\section{Appendices}

\subsection{HLC Algorithm}

Here, we provide the algorithm for HLC from \cite{KDMA2014OPODIS.HLC}. 

\begin{algorithm}
\algnewcommand\algorithmforsendandlocal{\textbf{Send/Local Event at process $i$}}
%\snote{Duong: Please change}
\algnewcommand\SENDORLOCALEVENT{\item[\algorithmforsendandlocal]}

\algnewcommand\algorithmforreceive{\textbf{Receive Event of message \em{m} at process $p_i$}}
\algnewcommand\RECEIVEEVENT{\item[\algorithmforreceive]}
\caption{Hybrid Logical Clocks (HLC) Algorithm from \cite{KDMA2014OPODIS.HLC}}
\label{alg:HLC}
\begin{algorithmic}[1] 
\SENDORLOCALEVENT
\State {$l$'$.i := l.i$} 
\State {$l.i := max(l$'$.i, pt.i)$ //tracking maximum time event, $pt.i$ is physical time at process $i$}
\State {If ($l.i = l$'$.i$) then $c.i := c.i + 1$  //tracking causality}
\State{Else $c.i := 0$}
%\State {Timestamp event {and/or message} with $l.a$,$c.a$}
\State {Timestamp the event (and the message for send event) with $l.i$, $c.i$}

\RECEIVEEVENT
\State {$l$'$.i := l.i$}
\State {$l.i := max(l$'$.i, l.m, pt.i)$ // $l.m$ is $l$ value in the timestamp of the message received}
\State {If ($l.i = l$'$.i = l.m$) then $c.i := max(c.i,c.m) + 1$}
\State {Elseif ($l.i = l$'$.i$) then $c.i := c.i + 1$}
\State {Elseif ($l.i = l.m$) then $c.i := c.m + 1$}
\State {Else $c.i := 0$}
\State {Timestamp event with $l.i$,$c.i$}
\end{algorithmic}
\end{algorithm}

\subsection{Omitted Proofs}

\subsubsection{Proof of Observation \ref{obs:syncclocks}}\hfill\newline
%\textbf{Proof. } \
This follows by induction and Observation \ref{obs:hptincrease}. Specifically, if $f$ is the a send event and $e$ is the previous event before $f$ then by Observation \ref{obs:hptincrease}, $\newclock.j + 1 \leq \cleanpt.j$. It follows that for event $f$, $\newclock.f$ would be set to be equal to $\cleanpt.j$. And, by definition of \lpt, $\lpt.f$ would be $0$. \QEDB
%\todo{does this need proof?}

\subsubsection{Proof of Lemma \ref{lem:lptfnonzeo}}\hfill\newline
%\textbf{Proof. } 
We consider two cases: Event $f$ is a local/send event (on process $j$) or it is a receive event (on process $j$). 
For the first, case, note that, by definition,  the least significant bits of \cleanpt.j are 0. If the value of $\lpt.j$ is set to $v, v > 0$, it must be due to the fact that $\newclock.f$ was set to $\newclock.j+1$, where $\newclock.j$ corresponded to the timestamp of the previous event on $j$. Thus, the above statement is satisfied by letting $e$ be the previous event on $j$.
%at that time, \lpt.j was $v-1$. Thus, the value of $l.e$ must be equal to $l.j$, where $l.j$ is the previous event on $j$. 
%
If $f$ is a receive event then a similar argument follows except that $e$ would be either the previous event on $j$ or the message send event. \QEDB

\subsubsection{Proof of Lemma \ref{lem:newclockbound}}\!\newline
%\textbf{Proof.}
%\dnote{(Do we need to consider the case where $f$ is the first event on the process?)}
The first part is trivially true.

For the second part, we can prove this by induction on the created events. The base case (initial events) is trivially satisfied. Next, let $e$ and $f$ be two consecutive events on process $j$. First, we consider the case where $f$ is a send event. By induction, we assume that $\newclock.e \leq \cleanpt.k_e + 2^{\ubits} $, where $\cleanpt.k_e$  was the value of $\cleanpt.k$ when $e$ was created. Let $\cleanpt.k_f$ be the value of $\cleanpt.k$ when $f$ was created. Clearly, $\cleanpt.k_e \leq \cleanpt.k_f$.
%\dnote{(By Observation~\ref{obs:hptincrease}, we have $\cleanpt.k_e \leq \cleanpt.k_f$)}. 
We consider two cases

\begin{itemize}
    \item $\newclock.f$ is set to $\newclock.e+1$. By induction, there exists $k$ such that $\newclock.e \leq \cleanpt.k_e + 2^\ubits$. Since $\cleanpt.k_e \leq \cleanpt.k_f$, we have $\newclock.e \leq \cleanpt.k_f + 2^\ubits$. Furthermore, if $\newclock.e = \cleanpt.k_f + 2^\ubits$ then creation of event $f$ would imply that Algorithm \ref{alg:timestamping} has an overflow. Hence, $\newclock.e < \cleanpt.k_f + 2^\ubits$. Thus, $\newclock.e+1 \leq \cleanpt.k_f + 2^\ubits$. In other words, $\newclock.f \leq \cleanpt.k_f + 2^\ubits$.

   % \dnote{$\newclock.f$ is set to $\newclock.e+1$. We have $\newclock.f = \newclock.e + 1 \leq \cleanpt.k_e + 2^{\ubits} + 1 \leq \cleanpt.k_f + 2^{\ubits}$. Hence, the condition is true for $f$}
    \item $\newclock.f$ is set to $\cleanpt.j$. Then, the statement is trivially true if we let $k=j$.
\end{itemize}

If $f$ is a receive event, the proof is similar except that we need to consider three cases for setting $\newclock.f$. \QEDB

\subsubsection{Proof of Theorem \ref{thm:boundeddiff}}\hfill\newline
\textbf{Proof. }
Without loss of generality, let $\newclock.j < \newclock.k$. Then, we have

\begin{tabbing}
\hspace*{1mm} \= \hspace*{5mm} \= 
$\newclock.k - \newclock.j$\\
\> $\leq$ \> $\newclock.k - \cleanpt.j$ \hspace*{15mm} \= // $\newclock.j \geq \cleanpt.j$ \\
\> $\leq$ \> $\cleanpt.k + 2^{\ubits} - \cleanpt.j $ \> // Lemma \ref{lem:newclockbound}\\
\> $\leq$ \> $pt.k + 2^{\ubits} - \cleanpt.j$ \> // $pt.k > \cleanpt.k$\\
\> $\leq$ \> $pt.k + 2^{\ubits} - (pt.j - 2^{\ubits})$ \> // $\cleanpt.j > pt.j - 2^{\ubits}$\\
\> $\leq$ \> $\epsilon + 2*2^{\ubits} = \epsilon + 2^{\ubits+1}$
\end{tabbing}

\begin{comment}
\subsection{Additional Figures}

\begin{figure}[h]
    \includegraphics[width=0.9\linewidth]{Images/BitsNeededBar.E6250.N8.S64.R.png}

    \caption{Log-scale number of events needing a given number of bits for the simulation of $\epsilon=6.25$ms$,$ $N=8,S=64$ in Random mode}
    \label{fig:BitsNeededBar.E6250.N8.S64.R}
\end{figure}

\end{comment}

\subsection{Experimental measurement of $\minsend, \minreceive$}
\label{sec:measure-minsend-minreceive}

We set up a cluster of machines connected via a switch in a local network. The switch capacity is 10 Gbps. Each machine is equipped with a 1000 Mbps network card. To measure the time to send a message $\delta_{send}$ (moving the message from the process to the network interface card), we have one machine send messages to 5 machines as fast as possible in 200 seconds. The value of $\delta_{send}$ is approximately calculated as the total time (200 seconds) divided by the number of messages sent by the sending machine. To measure $\delta_{receive}$ (moving the message from the network interface card to the receiving process), we have 5 machines sending messages to one machine in 200 seconds. The value of $\delta_{receive}$ is approximately calculated as the total time divided by the number of messages received at the receiving machine. All messages are sent via User Datagram Protocol. The message size is varied between 1 byte and 1,400 bytes.

From our experiments, we observe that the time to send or receive a message is a linear function of the message size. In particular:
$$\delta_{send}, \delta_{receive} = const_1 + const_2 \times message\_size$$

Where $const_1$ (unit is nanosecond) is the time to move the message through the protocol stack, $const_2$ (unit is nanosecond/byte) is the time it takes the network adapter to transmit (or receive, respectively) one byte of data to (or from, respectively) the cable.

The values of $const_1$ and $const_2$ depend on the machine and whether it is a receiver or a sender. In our measurement:
\begin{itemize}
    \item $const_1$ is a few microseconds. Specifically, it varies from 1,162 to 2,379 nanoseconds for senders and varies from 1,269 to 4,137 nanoseconds for receivers.
    \item $const_2$ varies from 6.6 to 7.6 nanosecond/byte for senders and varies from 6.1 to 15.5 nanosecond/byte for receivers. The advertised speed of network adapter on the machines in our lab is 1,000 Mbps (8 nanoseconds/byte).
\end{itemize}

With message payload between 1 byte and 1,400 bytes (to fit in an Ethernet frame), the time to send a message $\delta_{send}$ is between $1-12$ microseconds, and the time to receive a message $\delta_{receive}$ is between $1-24$ microseconds.

\end{document}